\documentclass[structabstract]{aa}
\usepackage{color}
\usepackage {amssymb}
\usepackage[english]{babel}
\usepackage{rotating}
\usepackage{longtable}
\usepackage{txfonts}
\usepackage{natbib}
\usepackage{pdfpages}

\newcommand{\kms}{km~s$^{-1}$ }

\newcommand{\one}{~{\sc i}}

\title{\object{RCW36}: characterizing the outcome of massive star formation\thanks{Based on observations performed with the ESO {\it New Technology Telescope} at La Silla Observatory, as part of program 074.C-0728, and with the ESO {\it Very Large Telescope} on Cerro Paranal, Chile, as part of programs 078.C-0780, 084.C-0604 and 087.C-0442.}}

\authorrunning{L.E. Ellerbroek et al.}
\titlerunning{RCW36: characterizing the outcome of massive star formation}

   \author{L.~E. Ellerbroek\inst{1}
          \and
          A. Bik\inst{2}
          \and
         L. Kaper\inst{1}
	\and
          K.~M. Maaskant\inst{3,1}
	\and
	M. Paalvast\inst{1}
	\and
	F. Tramper\inst{1}
	\and
	H. Sana\inst{1}
	\and
	L.~B.~F.~M. Waters\inst{4,1}
	\and
	Z. Balog\inst{2}
          }

   \institute{Astronomical Institute Anton Pannekoek, University of
     Amsterdam, P.O. Box 94249, 1090 GE Amsterdam,
     The Netherlands\\
              \email{l.e.ellerbroek@uva.nl}
         \and
Max-Planck-Institut f\"{u}r Astronomie, K\"{o}nigstuhl 17, Heidelberg, Germany
\and
Leiden Observatory, Leiden University, P.O. Box 9513, 2300 RA Leiden, The Netherlands  
         \and
SRON, Sorbonnelaan 2, 3584 CA Utrecht, The Netherlands
}
\date{today}

   \abstract{Massive stars play a dominant role in the process of clustered star formation, with their feedback into the molecular cloud through ionizing radiation, stellar winds and outflows. The formation process of massive stars is poorly constrained because of their scarcity, the short formation timescale and obscuration. By obtaining a census of the newly formed stellar population, the star formation history of the young cluster and the role of the massive stars within it can be unraveled.}
  {We aim to reconstruct the formation history of the young stellar population of the massive star-forming region RCW~36. We study several dozens of individual objects, both photometrically and spectroscopically, look for signs of multiple generations of young stars and investigate the role of the massive stars in this process.}
  {We obtain a census of the physical parameters and evolutionary status of the young stellar population. Using a combination of near-infrared photometry and spectroscopy we estimate ages and masses of individual objects. We identify the population of embedded young stellar objects (YSO) by their infrared colors and emission line spectra.}
  {RCW~36 harbors a stellar population of massive and intermediate-mass stars located around the center of the cluster. Class 0/I and II sources are found throughout the cluster. The central population has a median age of $1.1\pm0.6$~Myr. Of the stars which could be classified, the most massive ones are situated in the center of the cluster. The central cluster is surrounded by filamentary cloud structures; within these, some embedded and accreting YSOs are found.}
  {Our age determination is consistent with the filamentary structures having been shaped by the ionizing radiation and stellar winds of the central massive stars. The formation of a new generation of stars is ongoing, as demonstrated by the presence of embedded protostellar clumps, and two exposed protostellar jets.}


\begin{document}

\maketitle

\section{Introduction}

\begin{figure*}[!ht]
\centering
\includegraphics[width=0.54\textwidth]{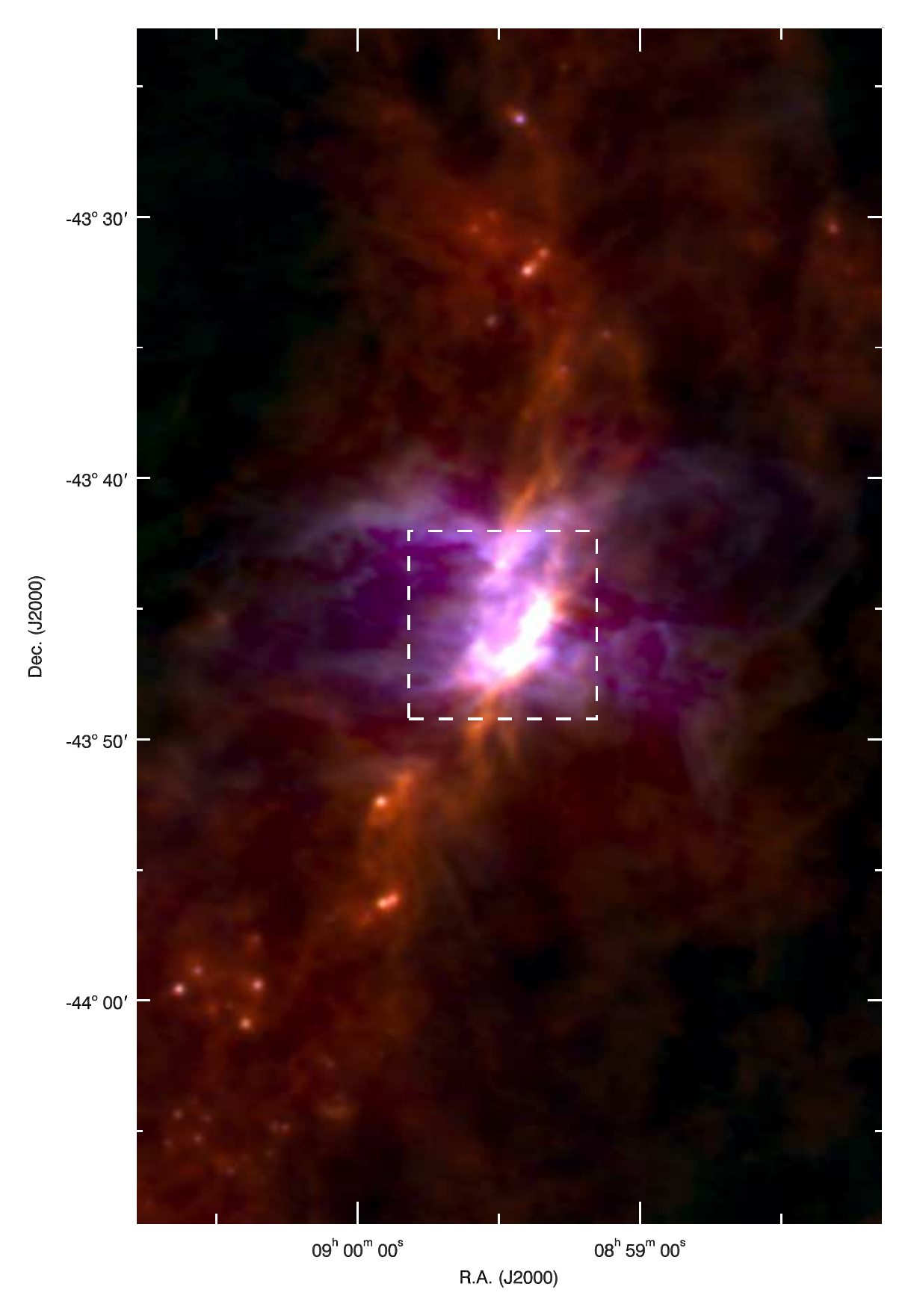}
\includegraphics[width=0.42\textwidth]{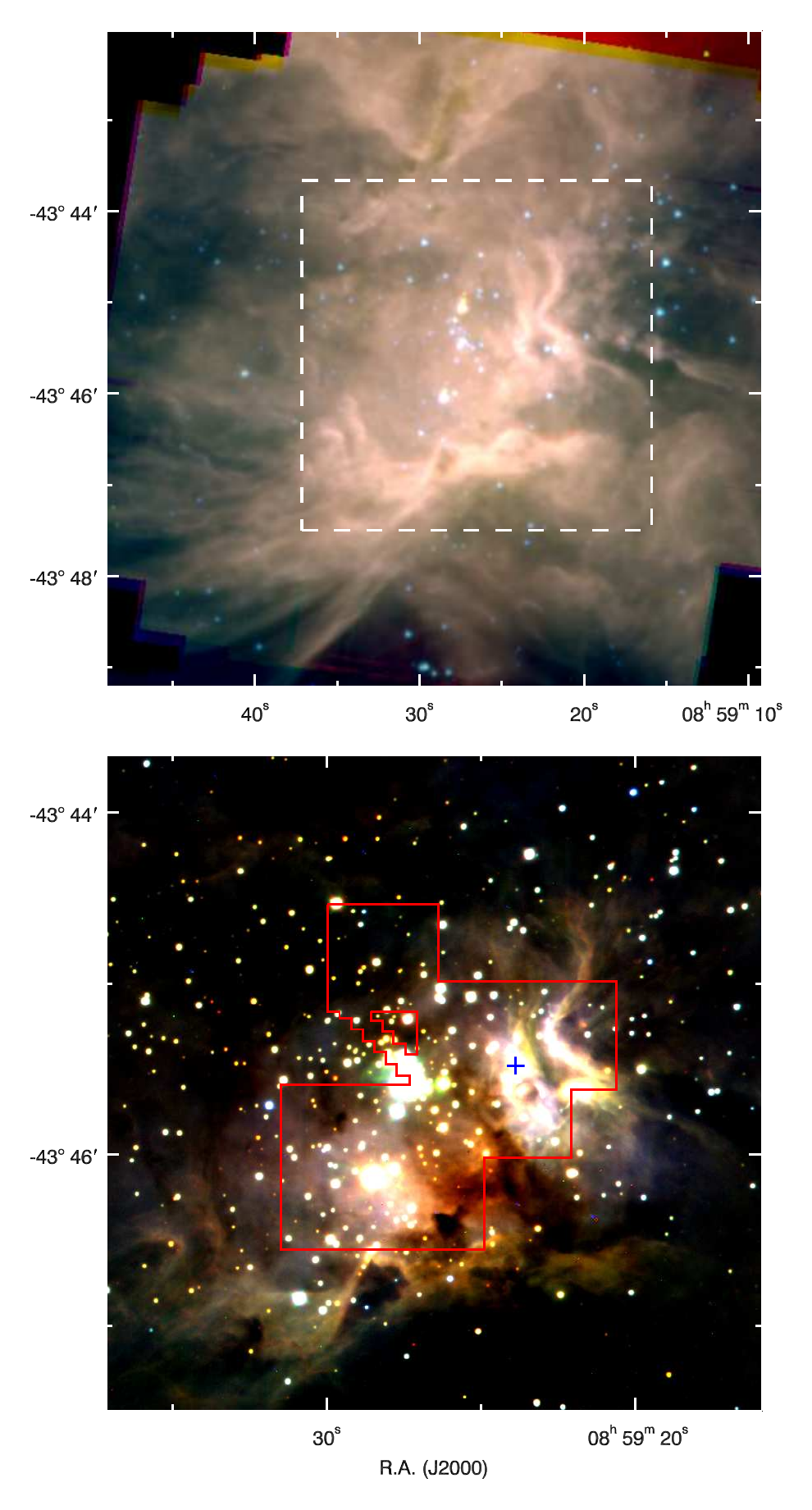}
\caption{\label{fig:overview} \textit{Left:} \textit{Herschel} image \citep{Hill2011} of RCW~36 and its surroundings (blue: PACS 70 $\mu$m, green: PACS 160 $\mu$m, red: SPIRE 250 $\mu$m). The cluster is located on a filamentary cloud structure that extends north to south; perpendicular to this ridge a bipolar nebula is seen. Note the ring-shaped structure in the center of this nebula. The dashed square denotes the region covered by \textit{Spitzer}. \textit{Top right:} \textit{Spitzer}/IRAC image of RCW~36 and its surroundings (blue: 3.6 $\mu$m, green: 4.5 + 5.8 $\mu$m, red: 8.0 $\mu$m, logarithmic scale). The dashed square denotes the region covered by NTT/SOFI. \textit{Bottom right:} NTT/SOFI broadband three-color image (blue: $J$, green: $H$, red: $K_{\rm s}$). The region delimited by the solid red lines is covered by the SINFONI observations. The blue cross denotes the peak location of the 5~GHz radio continuum source \citep{Walsh1998}. }
\end{figure*}

Most massive stars ($M > 10\,$M$_\odot$) form in clusters. As they evolve fast and are sources of ionizing radiation and stellar winds, they impact the evolution of their surrounding young stellar population and birth cloud. While massive stars are usually detected while on or already off the main sequence, their surrounding lower mass population is often still forming and provides a window, or ``clock'', on the star formation history. Therefore, a strategy towards understanding the complex process of clustered massive star formation is to study the outcome of star formation; to obtain a sample of star-forming regions, quantify the physical properties of the embedded young stellar population, and reconstruct the star formation history. With the combination of intermediate-resolution spectroscopy ($R \sim 10^3-10^4$) and multi-band photometry, it is possible to derive the stellar parameters of low- and high-mass stars and compare these to stellar evolution models to derive masses and ages. It also allows to characterize circumstellar material (disks and outflows) surrounding YSOs and, by virtue of the line-of-sight extinction, the local abundance of dust. Combining all these findings, a complete census of the young stellar population is obtained. Scenarios can then be conceived to describe the progression of star formation throughout the cluster, and the causes and effects of the formation of the massive stars within it. 

Over the past two decades, near-infrared imaging and spectroscopic surveys within the Galactic plane have revealed many young embedded stellar clusters showing a rich diversity in stellar content and evolutionary history \citep{Hanson1997, Walborn1997, Blum2000, Feigelson2008}. In a few of these studies, a massive star is identified as the source of ionizing radiation that formed the ultra-compact H~{\sc ii} region \citep[UCHII, e.g.][]{Watson1997, Alvarez2004, Bik2005}. In some cases, evidence of age spread and hence sequential star formation is found. In other cases, star formation seems to be triggered by the expansion of an H~{\sc ii} region \citep[e.g.][]{Zavagno2006, Zavagno2007}. However, it is generally difficult to derive a causal connection between the different generations of young stars. This is because evidence for triggered star formation is at best indirect, and because the uncertainties on stellar age estimates are large \citep[e.g.][]{Preibisch2012}. This is particularly the case for massive (proto)stars, as their position on the Hertzsprung-Russell diagram (HRD) strongly depends on their accretion history \citep{Davies2011}.

As part of the Formation and Early evolution of Massive Stars (FEMS) collaboration \citep{Bik2010}, we have obtained near-infrared images and spectra of several young embedded massive clusters, following up on a near-infrared survey of 45 southern star-forming regions centered on IRAS point sources exhibiting colors characteristic of UCHII regions (\citealt{Bik2004}; Kaper et al., in prep.). \citet{Bik2010} presented a spectroscopic census of RCW~34 in the Vela Molecular Ridge (VMR). They detected three distinct regions of star formation, suggesting that star formation progressed from south to north. \citet{Maaskant2011} studied the high-mass star-forming region GGD~12-15 centered on IRAS~06084-0611. They showed that the youngest generation of stars is centrally located, while somewhat more evolved objects are spread out over a larger area, suggesting sequential star formation along the line of sight. \citet{Wang2011} detected different evolutionary stages of star formation in the S255 complex. They conclude that their observations are best explained by the so-called triggered outside-in collapse star formation scenario, in which the filaments on the outskirts of the cluster collapse first, enhancing the instability of the massive star-forming cluster core.

The aim of this paper is to study the massive star-forming region RCW~36 (Gum 20, BBW~217) using a combination of photometry and spectroscopy that covers a broad range in wavelength ($0.3-8 \, \mu$m). Fig.~\ref{fig:overview} contains an overview of the observations. RCW~36 is located in cloud~C of the VMR, along a high-column density cloud filament which extends north to south \citep[NS, Fig.~\ref{fig:overview}, left;][]{Hill2011}. It includes a young star cluster (Massi et al. 2003) associated with the H~{\sc ii} region G265.151+1.454 of \citet{Caswell1987}. The region comprises the IRAS point source 08576-4334 with UCHII colors, also known as IRS 34 \citep{Liseau1992}, co-located with an UCHII region \citep{Walsh1998}. \citet{Hunt-Cunningham2002} suggest that star formation in RCW~36 is induced by a collision of two molecular gas clumps. These clumps are detected in several molecular emission lines at  different velocities north and south of the star-forming region. \citet[][hereafter MTHM13]{Minier2013} detect a tenuous, high-temperature bipolar nebula extending up to at least 10$'$ (2 pc) both east and west (EW) from the cluster. Around the origin of the nebula, a ring-like structure with high column density is detected; this structure is also visible in Fig.~\ref{fig:overview}. These authors propose a scenario where both the ring and the bipolar nebula are shaped (or ``blown out'') by the radiation pressure of the central O-star(s) in the cluster.

RCW~36 is very well suited for our scientific purposes, due to its relative proximity ($0.7$~kpc, \citealt{Liseau1992, Yamaguchi1999}), the presence of several O and B stars \citep{Bik2005} and YSOs \citep{Bik2006, Ellerbroek2011}. Based on observations in the far-infrared, \citet{Verma1994} estimate a total luminosity of $8\times10^4$~L$_\odot$, corresponding to the luminosity of two O9~V stars. This indicates that the emerging far-infrared luminosity is dominated by thermal reprocessing of radiation of the massive stars by dust. This confirms that the energy output from other massive (proto)stars embedded in the UCHII does not reach the observer, consistent with the high column density in this region \citep[$10^{23}$~cm$^{-2}$, $A_V \sim 100$~mag,][]{Hill2011}. 

Apart from the peaking far-infrared emission possibly indicating the presence of at least two (massive) protostellar cores \citep{Hill2011, Giannini2012}, a more evolved young stellar population is detected in the near-infrared. \citet{Baba2004} have performed near-infrared imaging in the $JHK$-bands and have detected more than 350 cluster members within the central $2 \farcm 5$ ($\sim 0.5$~pc). The same authors derive an age of 2--3~Myr and an average extinction of $A_V=8.1$~mag. \citet{Bik2005} classify two objects (objects 1 and 3 in this study, see Table~\ref{tab:dataoverview}) as O9~V\,--\,B1~V stars based on their K-band spectra, which is consistent with the result from \citet{Verma1994} and a distance of 0.7~kpc. 

\citet{Bik2006} report two near-infrared bright YSOs exhibiting Br$\gamma$ and CO emission (08576nr408 and \object{08576nr292}, our objects 2 and 4, respectively). \citet{BikThi2004} show that the CO emission from object 4 likely originates in a circumstellar Keplerian rotating disk. \citet{Ellerbroek2011} report the discovery of bipolar jets around two sources (HH~1042 and HH~1043, associated with 08576nr292 and \object{08576nr480}; our objects 4 and 97, respectively; see also \citealt{Ellerbroek2013}), adding to the evidence for ongoing star formation in this region. 

In this paper we perform a detailed analysis of the stellar content of RCW~36 using optical and near-infrared spectroscopy as well as near- and mid-infrared imaging. In Sect.~\ref{sec:obs} we describe the observations. Sect.~\ref{sec:photometry} identifies the stellar content of the region by the different photometric datasets; sources are assigned Lada classes \citep{Lada1987} according to their near- to mid-infrared spectral energy distributions (SED). We present detailed optical and near-infrared photospheric spectral classification of the pre-main sequence (PMS) population in Sect.~\ref{sec:spectroscopy}. In Sect.~\ref{sec:stellarpop} we combine all the results and refine the age estimate of the stellar population of RCW~36. We present a possible scenario for its star formation history, in which the massive and intermediate-mass stars have preceded a new generation of embedded protostars. Sect.~\ref{sec:conclusions} summarizes the conclusions of this work.


\section{Observations and data reduction}
\label{sec:obs}

To obtain a complete picture of the stellar content of RCW~36 we use archival near- to mid-infrared photometry to complement our near-infrared integral field VLT/SINFONI spectroscopy as well as optical to near-infrared VLT/X-shooter spectra of selected sources. The \textit{Herschel} images \citep[observed as part of the HOBYS program, observation ID 1342196658, P.I. Motte;][]{Hill2011}, which are used to show the large-scale structures with respect to the stellar population, were retrieved from the \textit{Herschel} Science Archive.

\subsection{Near-infrared imaging and photometry: NTT/SOFI}
\label{sec:obs:sofi}

We have retrieved broad-band near-infrared $JHK_{\rm{s}}$ data from the ESO archive. The observations (P.I. Bialetsky) were carried out with SOFI \citep{Moorwood98} on the \emph{New Technology Telescope} (NTT) at La Silla Observatory in Chile. The observations were performed on 18 May 2005 under decent seeing conditions (0.9$''$ in $K_{\rm{s}}$, 0.8$''$ in $H$ and 1.2$''$ in the $J$-band). The \emph{Detector Integration Time} (DIT) was 10, 8, and 6 seconds for $J$, $H$, and $K_{\rm{s}}$, NDIT~=~12 and a total number of 6 frames were taken on source. This results in a total exposure time of 12 min ($J$), 9.6 min ($H$), and 7.2 min ($K_{\rm{s}}$) for the three bands. Offset sky positions were taken to ensure a good sky subtraction.

The data were reduced using the ESO pipeline (version 1.5.2) for  SOFI. We corrected the frames with darks and flat fields which were obtained on the same morning as the science observations. After that the data were sky subtracted and the final mosaic was created. We obtained an astrometrical solution by matching the positions of the stars with those of 2MASS \citep{Skrutskie06}. 

Photometry was performed using \emph{daophot} \citep{Stetson87} under the IRAF environment. First, stellar sources were detected using the \emph{daofind} task and aperture photometry was performed using the task \emph{phot} with an aperture equal to the Full With Half Maximum (FWHM) of the stellar sources. Using the tasks \emph{pstselect} and \emph{pst} a reference point spread function (PSF) was constructed by using over 25 bright, isolated stars in the SOFI images. Finally, PSF-fitting photometry was performed using the task \emph{allstar} on all the sources detected with a 3$\sigma$ threshold. The absolute calibration of the photometry was performed by comparing the photometry of bright, isolated stars with their 2MASS values. No significant color terms were found in the photometric calibration.

The stars with $K_{\rm{s}} <$~10.8~mag (20 stars) are saturated in the 2005 SOFI images. For those stars the 2MASS photometry is used instead. For two objects (3 and 7 in Table~\ref{tab:dataoverview}), the 2MASS photometry is contaminated by a neighboring bright star and the SOFI $J$ and $K_{\rm s}$ band magnitudes of \citet{Bik2004} are used. The magnitudes of these objects in the $H$-band (which was not covered by \citealt{Bik2004}) are calculated by performing spectrophotometry on their SINFONI spectra. The seeing conditions during these observations were sufficient to avoid contamination by neighboring stars. The SINFONI $K$-band values of these stars agree well with the \citet{Bik2004} observations.

The limiting magnitudes at 10$\sigma$ are approximately 19.1, 19.3, and 18.4~mag in $J$, $H$, and $K_{\rm s}$, respectively. A total of 745 sources are detected in the $K_{\rm s}$-band with a photometric error of $<0.1$~mag. Of these, 395 are detected in all three bands with a photometric error of $<0.1$~mag and a positional agreement of $< 0\,\farcs 5$. The point sources are numbered according to their $K_{\rm s}$ magnitude.

\subsection{Mid-infrared imaging and photometry: Spitzer/IRAC}
\label{sec:obs:spitzer}

Imaging data taken with IRAC \citep{Fazio04} on board of \emph{Spitzer} have been retrieved from the \emph{Spitzer} archive (program ID 20819, P.I. Tsujimoto). The data of RCW~36 have been taken in the high-dynamic range mode, consisting of a set of deep images with a frame time of 10.4~s and a set of images taken with a frame time of 0.4~s to ensure that the brightest sources were not saturated. The raw data have been processed with the standard IRAC pipeline (version 18.18.0) to create the \emph{basic calibration data} (BCD). These BCD were downloaded from the \emph{Spitzer} archive and processed by custom IDL routines as described in \citet{Balog07}.

As the point spread function of the IRAC images is undersampled, obtaining aperture photometry is preferred. Aperture photometry on the reduced mosaic was performed using \emph{daophot} inside IRAF. The sources (618 in band 1) were detected using \emph{daofind} and photometry was performed using \emph{phot} with an aperture equal to the FWHM of the stellar PSF.  The background is measured in an annulus between 2 and 6 pixels around the star. The photometry is corrected with the aperture corrections taken from the IRAC handbook. Sources with a positional agreement of 2 pixels ($2\farcs4$) between the different IRAC bands were matched. Some matches were discarded because of the faulty detection of some of the filaments in the 8~$\mu$m IRAC-band as point sources.

The stellar population under study is located on the forefront of a dense dust cloud with $A_V$ up to 100 mag \citep{Hill2011}. Contamination from extragalactic background sources is thus expected to be negligible. Moreover, only two sources dimmer than the brightness limit formulated by \citet{Gutermuth2009} for extragalactic sources (i.e. [3.6~$\mu$m] $> 15$~mag) were detected.

IRAC detections were matched with SOFI sources within $0 \farcs 5$. Increasing this radius with a factor 2 did not lead to a different result. Some spurious matches were discarded upon careful examination of the images. Only the sources with a photometric error of less than 0.1 mag (0.2 mag for IRAC) are used for the analysis. This results in 250 matches between IRAC band 1 and SOFI $K_{\rm s}$ and 23 matches between all IRAC bands and SOFI $K_{\rm s}$. The limiting magnitudes at 10$\sigma$ were 15.3, 12.7, 11.4 and 7.5~mag in bands 1, 2, 3 and 4 respectively.

\subsection{Near-infrared integral field spectroscopy: VLT/SINFONI}
\label{sec:obs:sinfoni}
Near-infrared $H$- and $K$-band spectra have been taken with the integral field spectrograph SINFONI \citep{Eisenhauer2003,Bonnet2004}, mounted on UT4 of the ESO \textit{Very Large Telescope} (VLT) on Cerro Paranal in Chile. The data were obtained in service mode between February 28 and March 23, 2007, with a typical seeing of $0\farcs8$.  RCW~36 was observed using the non-adaptive optics mode with the $0\farcs250$ pixel scale, resulting in an $8'' \times 8''$ field of view. To obtain an $H$- and $K$-band spectrum the $H+K$ grating was selected resulting in a spectral resolution of $R\sim1500$ and a wavelength coverage from $1.55-1.75\, \mu$m and $2.00-2.50\, \mu$m.  

To cover the area of the cluster as shown in Fig.~\ref{fig:overview}, a mapping pattern was applied with offsets of $\Delta \alpha = 4\farcs00$ and $\Delta \delta = 6\farcs75$. The offsets were designed such that every pixel in the field of view is covered as least twice. A detector integration time (DIT) of 30 seconds per integration was chosen. Sky frames were taken every 3 minutes on carefully selected offset positions with the same integration times to ensure an accurate sky subtraction. After every science observation a telluric standard star was observed at the same airmass to enable correction for the telluric absorption lines.

The SINFONI data are reduced using the SPRED software package version 1.37 \citep{Schreiber2004, Abuter2006}. The data reduction procedure is described in detail in \citet{Bik2010} and consists of dark and flat field correction of the raw data. After a distortion correction, the merged 3D datacubes were created. Telluric standard stars were used to correct for the telluric absorption lines, as is described in \citet{Ellerbroek2011}.

\subsection{Optical to near-infrared spectroscopy: VLT/X-SHOOTER}
\label{sec:obs:xshooter}

Eight objects have been observed with X-shooter, mounted on UT2 on the VLT, resulting in optical to near-infrared ($300 - 2500$~nm) spectra, see Table~\ref{tab:xshobs}. The slits used were $1\farcs 0$ (UVB, $300 - 590$~nm), $0\farcs 9$ (VIS, $550 - 1020$~nm) and $0\farcs 4$ (NIR, $1000 - 2480$~nm). This resulted in a spectral resolution of 5000, 9000 and 11,000 in the three arms, respectively. Directly before or after these observations the A0V star HD80055 was observed in order to remove telluric absorption lines in the near-infrared. A spectrophotometric standard was observed each night for flux-calibration. The spectra were reduced using the X-shooter pipeline \citep[version 1.3.7][]{Modigliani2010}.

The X-shooter spectra of the early-type stars (1, 2, 3 and 10) ensure a more precise spectral classification than that obtained with SINFONI, while the spectrum of object 2 also contains information on its circumstellar material. This is also the case for the young stellar objects (4, 9 and 97). Finally, the late-type PMS star 26 was observed with X-shooter in order to check the consistency of the spectral classification of the SINFONI spectra of late-type stars.

\begin{table}
\begin{minipage}[c]{\columnwidth}
    \renewcommand{\footnoterule}{}

\centering
\caption{\normalsize{Journal of X-shooter observations.}}
\renewcommand{\arraystretch}{1.4}
\begin{tabular}{llllll}
\hline 
\hline
\multicolumn{2}{c}{Object}  & HJD & Exp. time & \multicolumn{2}{c}{continuum S/N} \\
\# & B05\footnote{See \citet{Bik2005, Bik2006, Ellerbroek2013}.} & & (s) & 460~nm & 800~nm \\
\hline
1 & 462 & 22, 23-02-2010 & 2400 & 35 & 86\\
2 & 408 & 23-02-2010 & 2400 & 14 & 88 \\
3 & 413 & 22, 23-02-2010 & 2400 & 29 & 83\\
4 & 292 & 22, 23-02-2010, & 6000 & 9 & 58\\
 & & 18-01-2011 & & &\\
9 &   & 22-04-2011 & 1800 & $<3$ & 24 \\
10 & 179 & 23-02-2010 & 3600 & 4 & 87 \\
26 & & 12-02-2011 & 1200 & $<3$ & 13 \\
97 & 480 & 12-02-2011 & 1800 & $<3$ & $<3$ \\
\hline
\vspace{-15pt}
\end{tabular} 
\label{tab:xshobs}

\end{minipage}
\end{table}

\section{Results from photometry}
\label{sec:photometry}

In this section, we use the near-infrared SOFI photometry to identify the stellar population (Sect.~\ref{sec:photometry:sofi}) and the IRAC photometry to reveal the circumstellar material surrounding it. The stellar density increases towards the location of the massive stars, objects 1 and 3 (see Fig.~\ref{fig:overview}), which we define as the center of the cluster.

\begin{figure}[!ht]
   \centering
\includegraphics[width=0.44\textwidth]{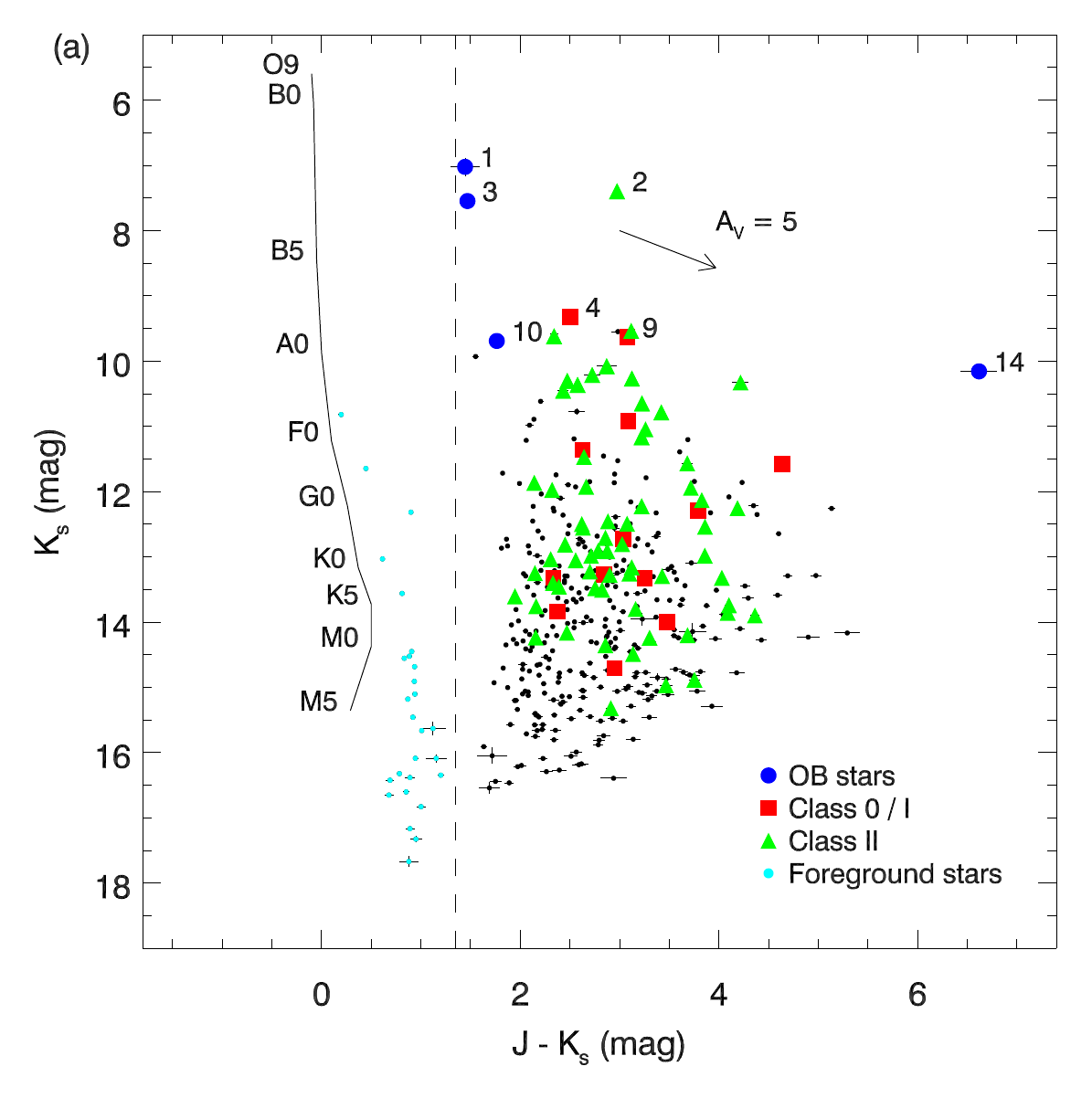} 
\includegraphics[width=0.45\textwidth]{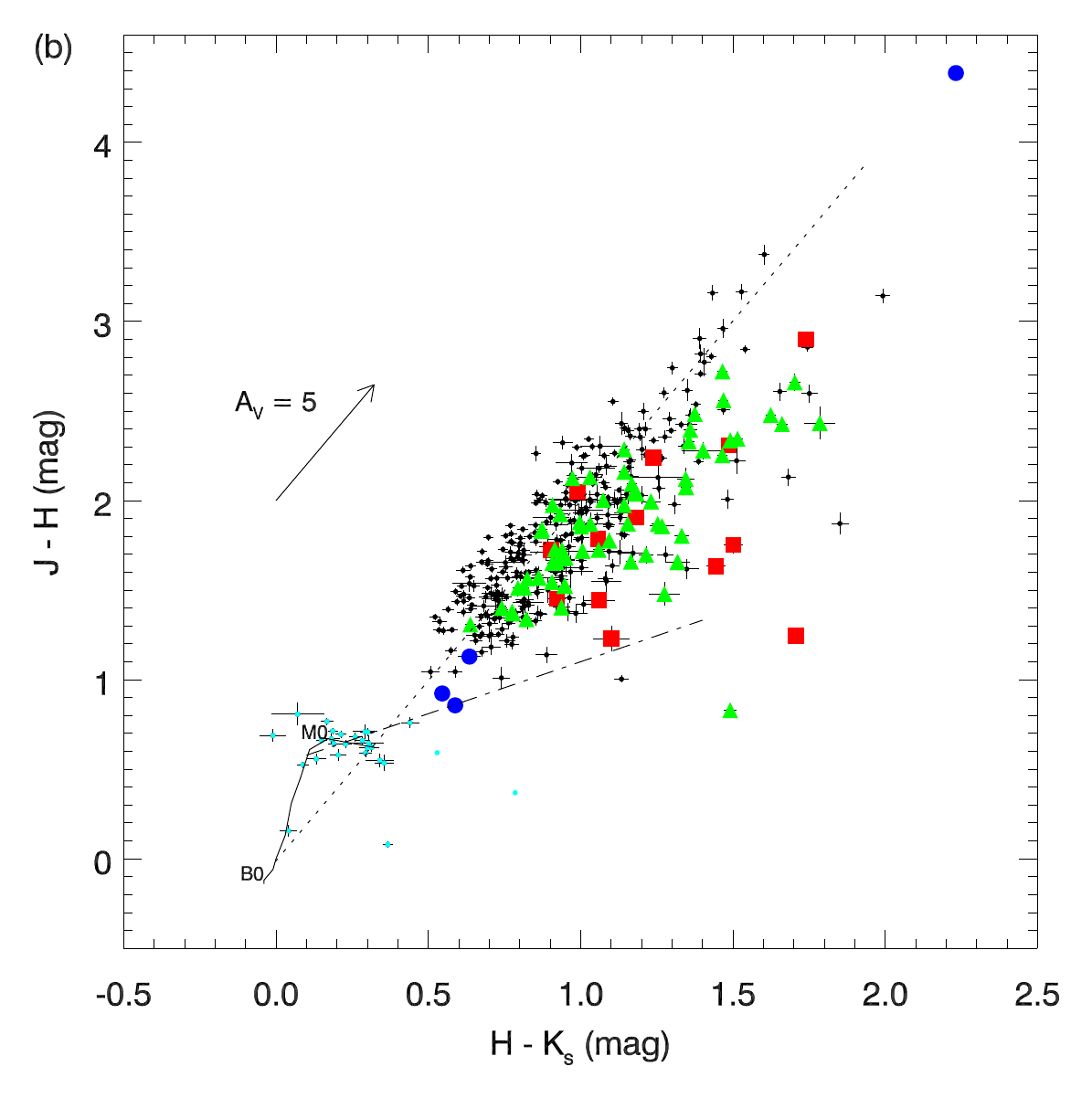}
   \caption{(\textit{a}) ($J-K_{\rm s}, K_{\rm s}$) color-magnitude diagram. (\textit{b}) ($H-K_{\rm s}$, $J-H$) color-color diagram. All detections in the SOFI-field of view (Fig.~\ref{fig:overview}, bottom right) are included, except for those with an error larger than 0.1 mag in any of the color indices. Blue symbols indicate OB stars (see Sect.~\ref{sec:spectroscopy:et}); red (Class 0/I) and green (Class II) symbols indicate Lada classified sources (see Sect.~\ref{sec:photometry:irac}). The solid line is the zero-age main sequence  \citep[ZAMS,][]{Blum2000, Bessell1988} at a distance of 0.7~kpc. The dashed line indicates the separation between the cluster and the foreground population, which is indicated with cyan symbols. The dotted line is the extinction vector \citep{Cardelli1989} and the dot-dashed line is the cTTS locus \citep{Meyer1997}.
   }
   \label{fig:soficolors}
\end{figure}
\begin{figure}[!t]
\centering
\includegraphics[width=0.44\textwidth]{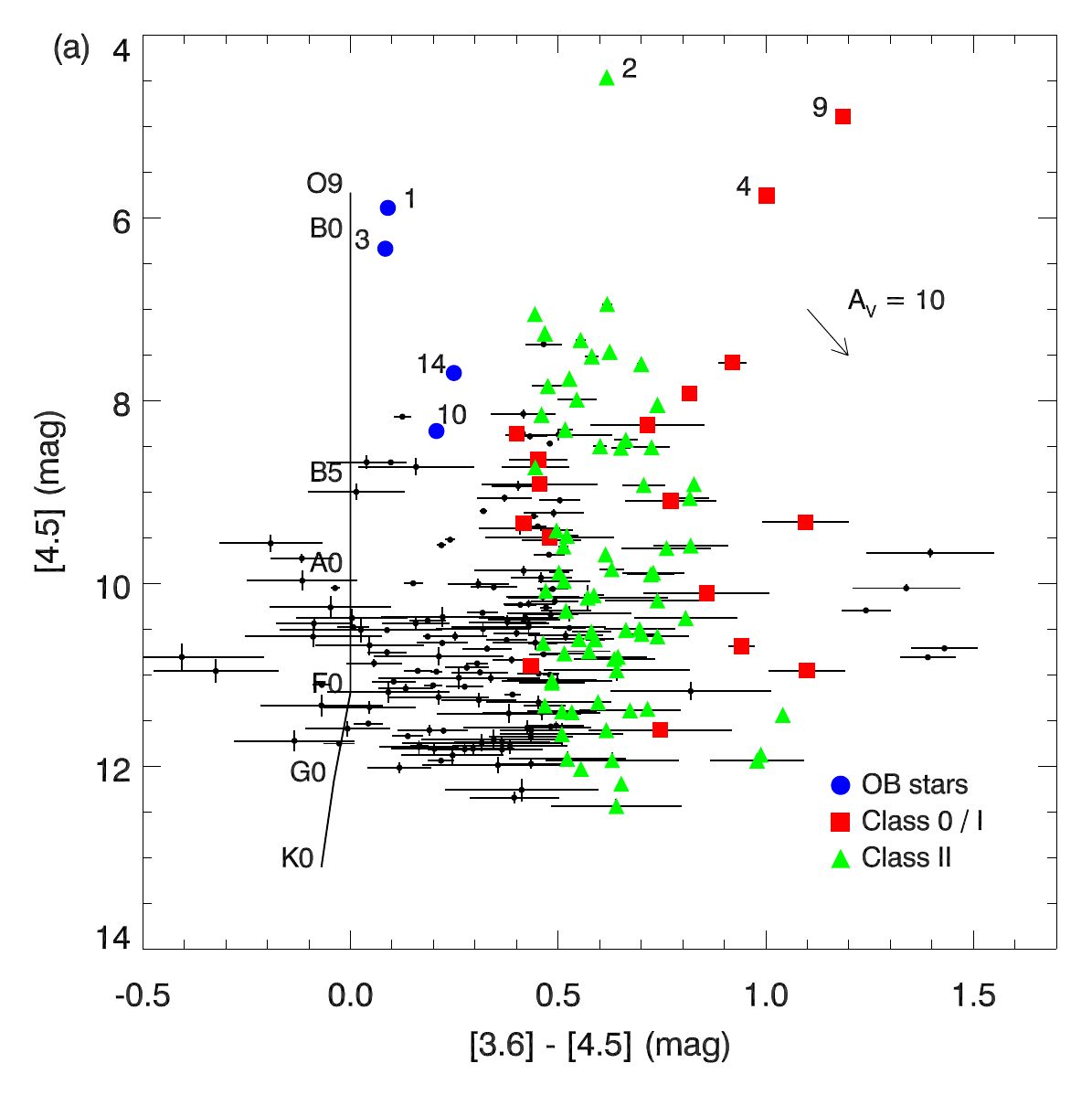} 
\includegraphics[width=0.45\textwidth]{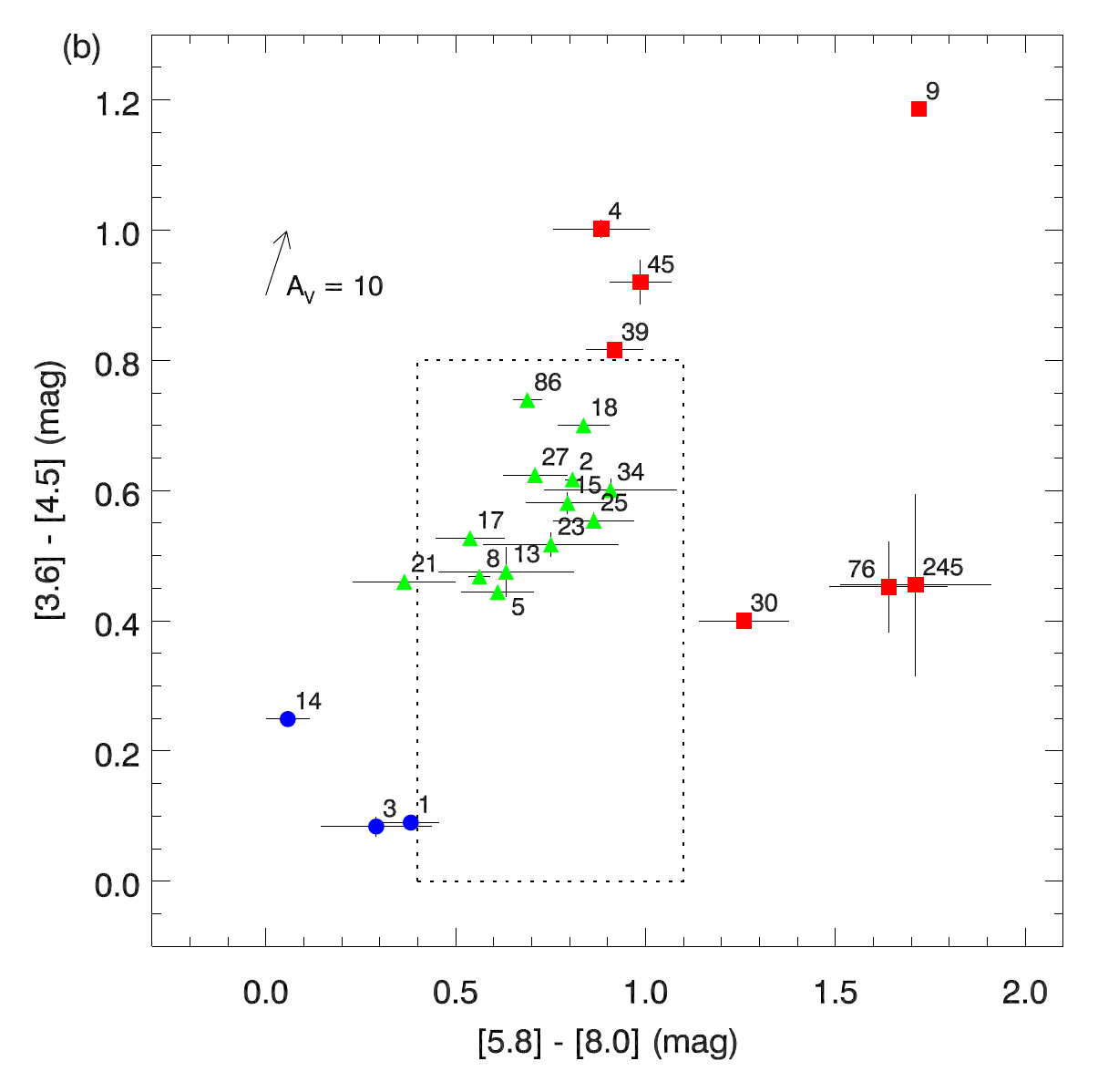} 
\caption{\label{fig:spitzercolors} (\textit{a}) IRAC color-magnitude diagram with the sources detected in the first two IRAC bands. Five sources which are also observed with X-shooter are labeled; plot symbols are the same as in Fig.~\ref{fig:soficolors}. (\textit{b}) IRAC color-color diagram with the sources within the SOFI field of view detected in all four IRAC bands. The dotted square indicates the location of the class II sources \citep{Megeath2004}.}
\end{figure}
\subsection{Near-infrared imaging}
\label{sec:photometry:sofi}

The ($J-K_{\rm s}, K_{\rm s}$) color-magnitude diagram (Fig.~\ref{fig:soficolors}a) shows a reddened stellar population consistent with a distance of 0.7~kpc. The population blueward of $(J-K_{\rm s})=1.35$ mag probably consists of low-mass foreground stars and is not considered to be part of the cluster. 

Fig.~\ref{fig:soficolors}b shows the ($H-K_{\rm s}$, $J-H$) color-color diagram. Also here the foreground population is clearly visible. To correct for reddening due to interstellar extinction. We adopt the extinction law from \citet{Cardelli1989} with the total-to-selective extinction parameter $R_V$ set to the average Galactic value of $3.1$ (which may be an underestimate, see also Sec.~\ref{sec:spectroscopy:nebular}). We conclude that the majority of the sources are found along the reddened location of the main sequence. However, several objects are located below this reddening line. These objects possess a near-infrared excess, indicative of a circumstellar disk and hence of their young age.

Using the above mentioned extinction law, we calculated the average extinction towards the stellar population of RCW~36. Excluding the foreground population, the average reddening detected towards the main sequence is $A_V=14.7$ mag with a 1$\sigma$ spread of 5.5 mag. This large spread on the average value of the extinction suggests that differential extinction is strongly affecting the appearance of the stellar population.

An alternative way to estimate the average interstellar extinction is obtained by dereddening to the locus of the classical T Tauri stars \citep[cTTS,][]{Meyer1997}; this assumes for every source an intrinsic infrared excess due to a circumstellar disk. This likely leads to a more accurate estimate of $A_V$. The mean extinction obtained by dereddening all sources to the cTTS locus is $A_V = 10.1 \pm 4.6$~mag, about 30\% less than the aforementioned value, but consistent within the uncertainty. This estimate agrees with the average value of $A_V = 8.1$~mag found by \citet{Baba2004} who use the same method. 

The photometric data of RCW~36 do not show distinct subgroups within the stellar population that have a different extinction or infrared colors, apart from the ``foreground population'' defined above. For a more thorough treatment of the extinction properties, see Sect.~\ref{sec:spectroscopy:nebular}.

\begin{figure*}
\centering
\includegraphics[height=0.4\textwidth]{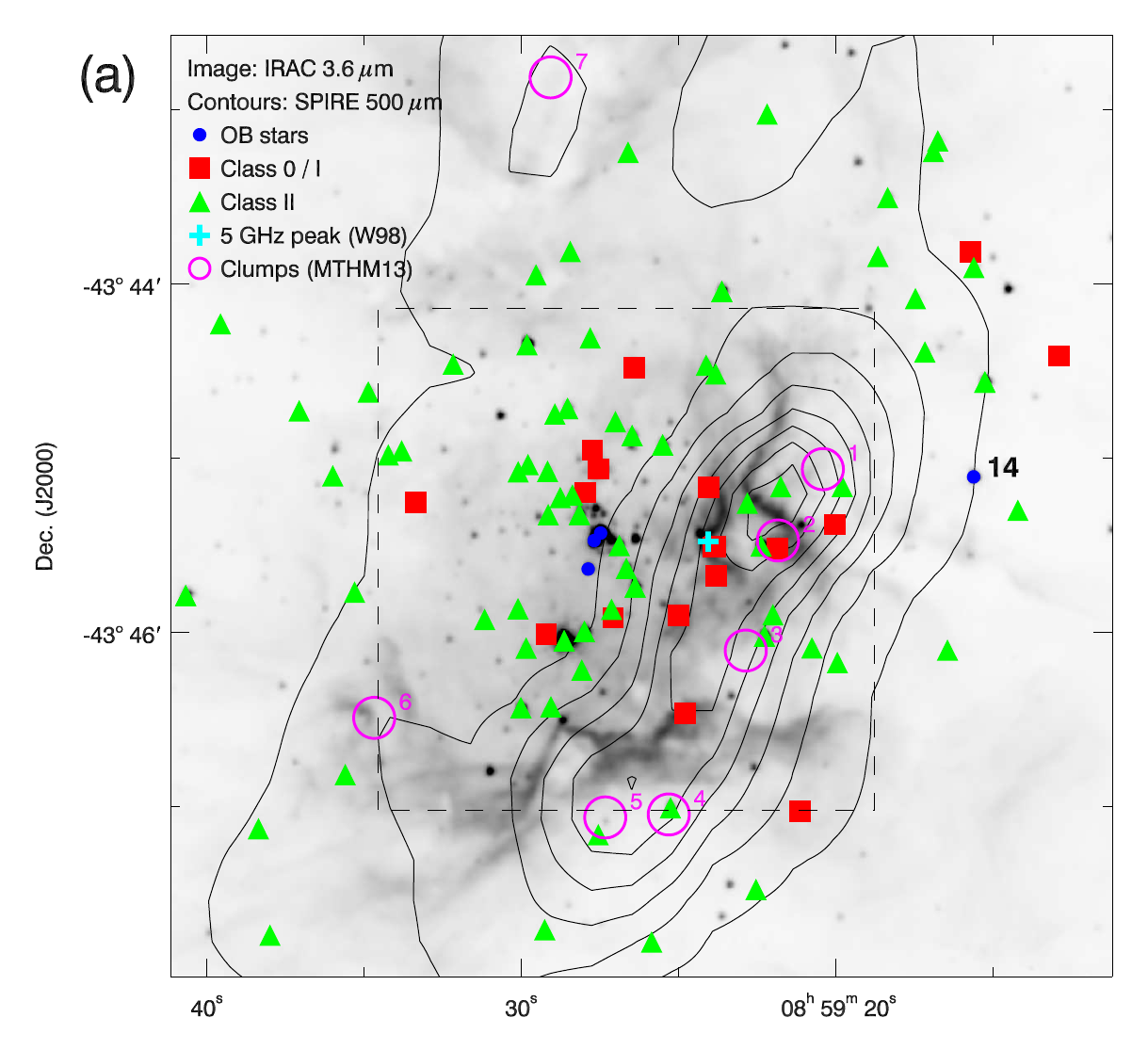} 
\includegraphics[height=0.4\textwidth]{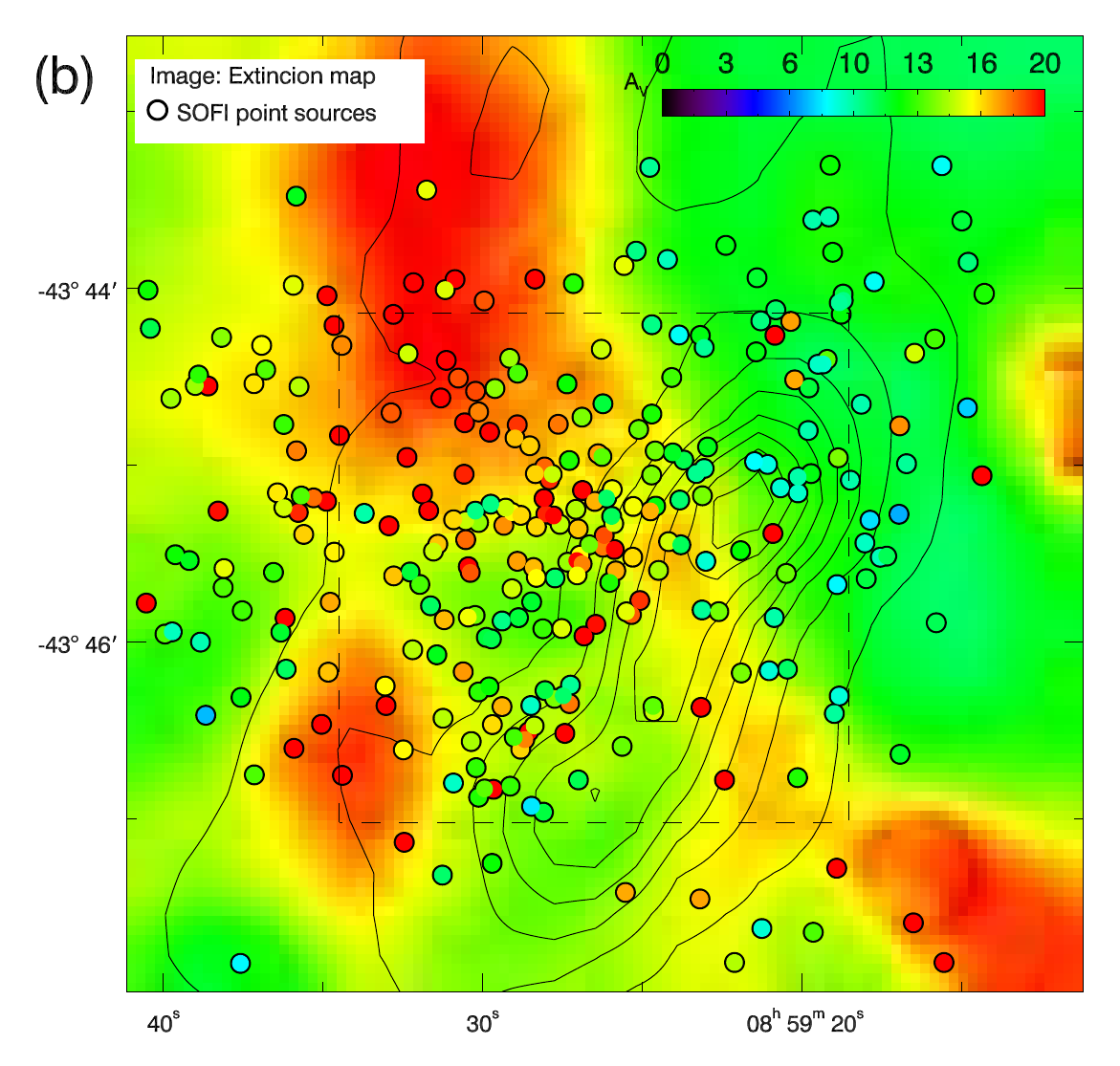} \\
\includegraphics[height=0.4\textwidth]{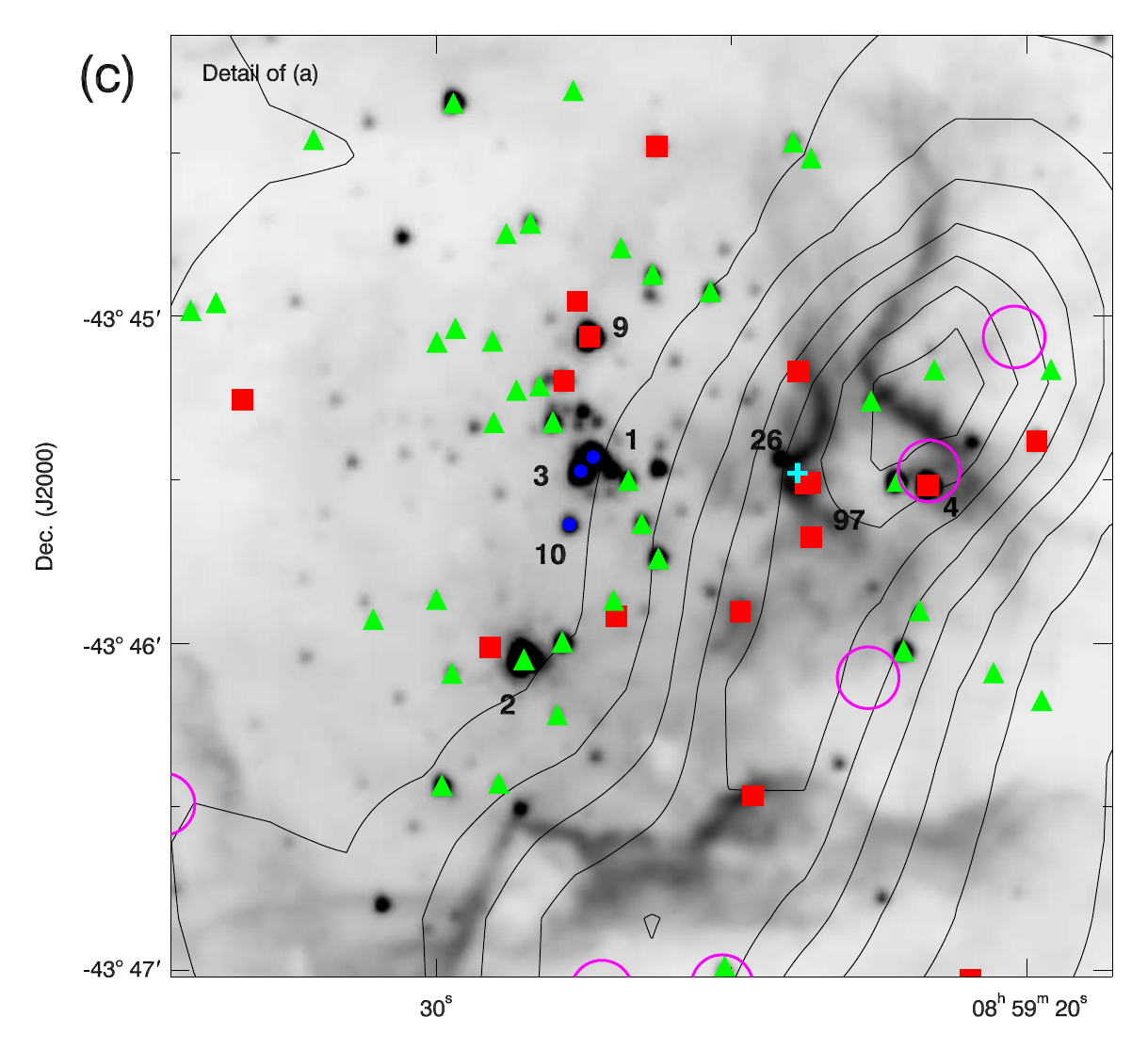} 
\includegraphics[height=0.4\textwidth]{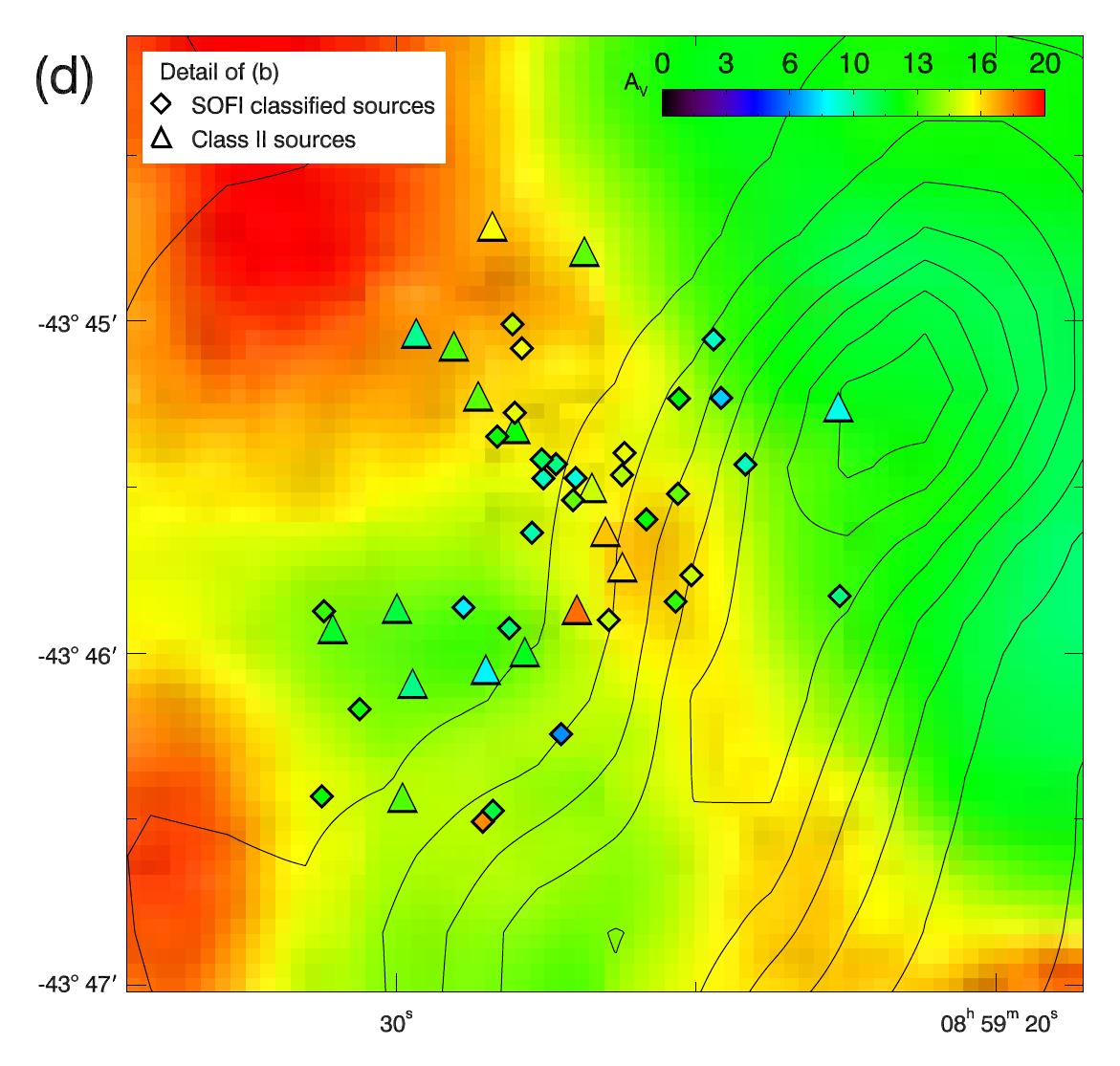} \\
\includegraphics[height=0.416\textwidth]{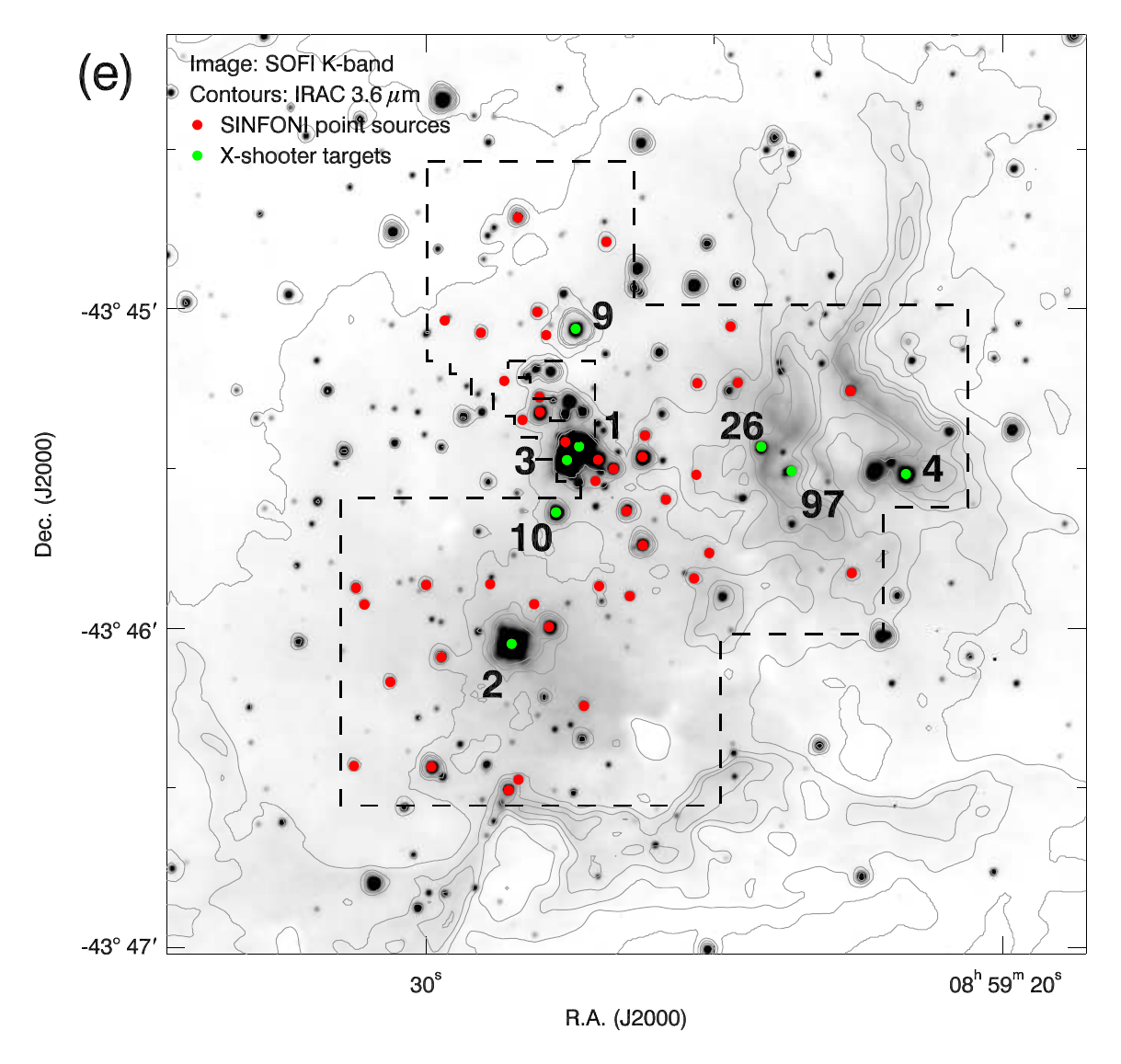}
\includegraphics[height=0.416\textwidth]{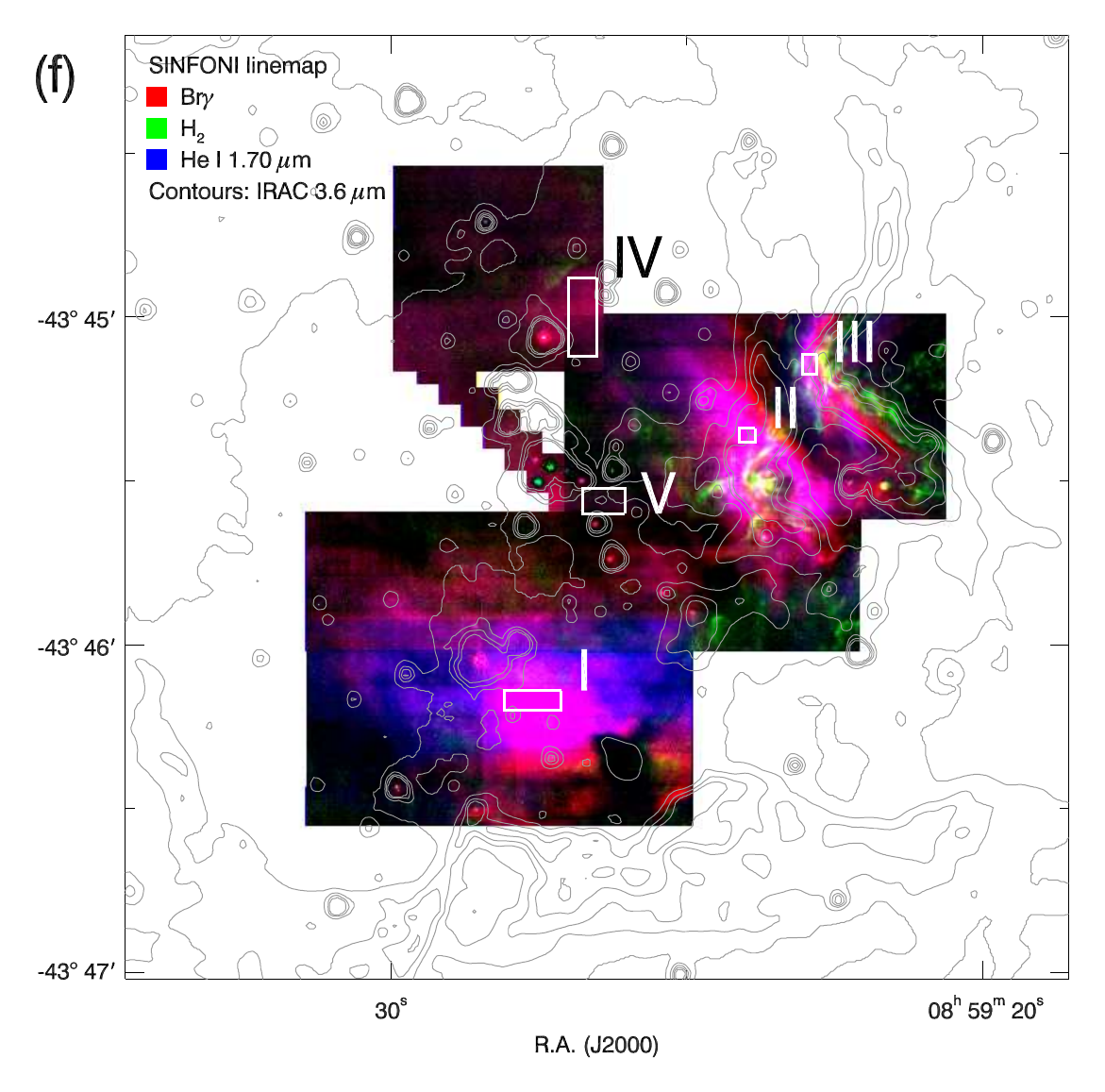} \\
\caption{\label{fig:maps} 
(\textit{a}) Map of RCW~36; the same symbols as in Fig.~\ref{fig:soficolors} are used to classify the stellar population. Purple circles are the protostellar cores from MTHM13; according to these authors, core 1 contains over 20 M$_\odot$. The cyan cross marks the location of the UCHII \citep[][W98]{Walsh1998}. The SPIRE 500~$\mu$m contours are overplotted.
(\textit{b}) Extinction map based on the $JHK_{\rm s}$ colors of individual stars (see text). The SOFI point sources are color-coded with their extinction values, overlain on the spatially smoothed extinction map.
(\textit{c}) Same as (\textit{a}), zoom in on boxed region.
(\textit{d}) Same as (\textit{b}), zoom in on boxed region. Symbols indicate the classified stars, color-coded with their extinction values obtained by spectral classification. 
(\textit{e}) Map showing the targets of which a spectrum was obtained and classified. 
(\textit{f}) SINFONI three-color linemap showing the different ionization properties of the nebula. Regions I--V (see text) are indicated.
The symbols used in (\textit{a}), (\textit{b}) represent all classified sources; in (\textit{c}) all SOFI detections and (\textit{d}) only sources with a spectral classification.
Figs. (\textit{a})--(\textit{b}) and (\textit{c})--(\textit{f}) have the same scale, respectively.
}
\end{figure*}

\begin{table*}[!ht]
\caption{\normalsize{Stellar parameters from optical spectra.}}
\begin{minipage}[c]{\textwidth}
    \renewcommand{\footnoterule}{}
\renewcommand{\arraystretch}{1.4}
\begin{center}
\begin{tabular}{llllllll}
\hline 
\hline
Object (old id\footnote{See \citet{Bik2005, Bik2006}.}) & $T_{\rm eff}$ ($10^3$ K) & $\log g$ & $v_{\rm rot}$ (km~s$^{-1}$) & Sp. type\footnote{From optical diagnostics and reference spectra, see text.} & $R_*$ (R$_\odot$) & $A_V$ (mag) & $d$ (kpc)\footnote{From spectroscopic parallax.}\\
\hline
1 (462) & $34.2^{+1.3}_{-2.6} $ & $4.22^{+0.26}_{-0.24}$ & $160^{+36}_{-38}$ & O8.5--9.5 V &  $8.2\pm 0.6$ & $10.4\pm0.5$  & $0.67\pm 0.17$\\
3 (413) & $34.9^{+1.6}_{-2.2} $   & $4.09^{+0.18}_{-0.29}$ & $34^{+24}_{-24}$ & O9.5--B0 V &  $6.0\pm0.2$  & $9.6\pm0.5$ & $0.81\pm0.19$  \\
10 (179) &  $19.8^{+1.4}_{-2.1} $ & $3.5-4.0$ & $148^{+36}_{-38}$ & B2--3.5 IV &  $2.9\pm0.3$  & $9.8\pm0.5$ & $0.62\pm0.16$ \\
2 (408) & $8.72^{+0.25}_{-0.26} $  & $3.5-4.0$ & \dots & A2--4 IV  &  $5.1\pm0.5$ & $8.5\pm0.5$ & \dots \\ 
\hline
    \vspace{-18pt}
\end{tabular}
\end{center}
\end{minipage}
\label{tab:ostars}
\end{table*}

\begin{figure*}
\centering
\includegraphics[width=\textwidth]{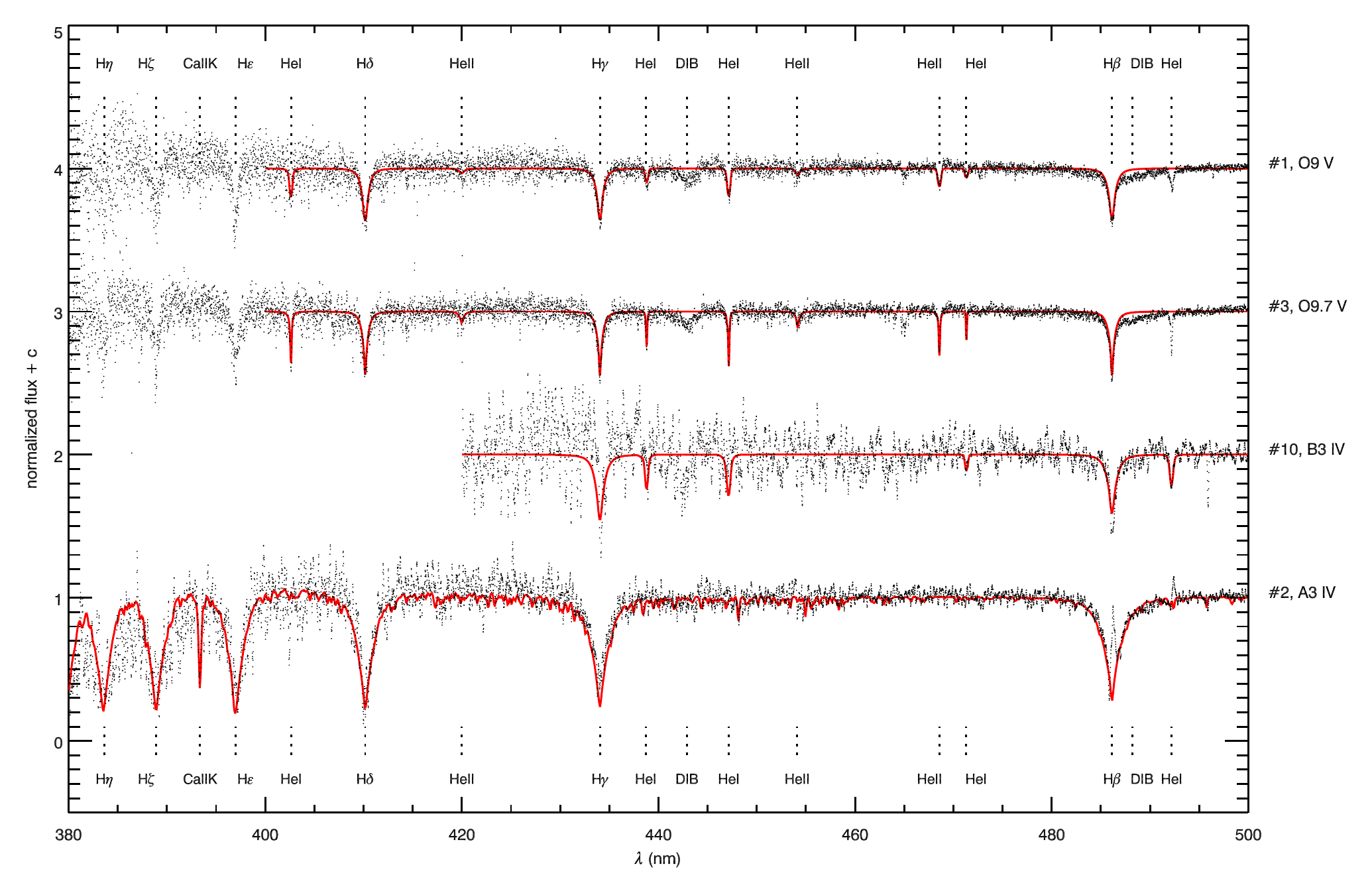} 
\caption{\label{fig:etspectra} Optical spectra of the four classified early-type stars in RCW~36. The spectra of objects 2 and 10 are rebinned with 3 pixels. Overplotted in red are the best-fit model spectra: synthetic spectra from FASTWIND (objects 1, 3 and 10) and \citet[][object 2]{Munari2005}. Only fitted profiles are shown for objects 1, 3 and 10. Note the emission component in the H$\beta$ line of object 2, which probably originates in a disk or wind.}
\end{figure*}

\subsection{Mid-infrared imaging}
\label{sec:photometry:irac}

Fig.~\ref{fig:spitzercolors}a displays the IRAC colors of the 210 sources which were detected (with a photometric error $<0.2$ mag) in the first two IRAC bands and with a $K_{\rm s}$-band counterpart. Using the classification scheme of \citet{Gutermuth2009}, which makes use of the $JHK_{\rm s}$ and first two IRAC bands, we classified 18 of these sources as class 0/I and 70 sources as class II. This classification is consistent with the classification scheme of \citet{Megeath2004} and \citet{Allen2004} for the objects that are detected in all four IRAC bands (Fig.~\ref{fig:spitzercolors}b). We find that about half of the sources detected in $H$, $K_{\rm s}$ and at least in the first two 2 IRAC bands also have an intrinsic infrared excess. This suggests a high disk fraction in this cluster and hence a young age \citep{Lada2003, Hernandez2008}. A more accurate estimate of the disk fraction would be obtained by a completeness analysis, which is beyond the scope of this paper.

The spatial distribution of the Lada classified sources (Fig.~\ref{fig:maps}a,\,c) shows that many of the class 0/I sources are associated with the filamentary structures. Although they trace point sources in all IRAC bands, their flux in band 4 is possibly contaminated by emission from the filaments, resulting in a classification as class 0/I.

\begin{figure*}
\centering
\includegraphics[width=\textwidth]{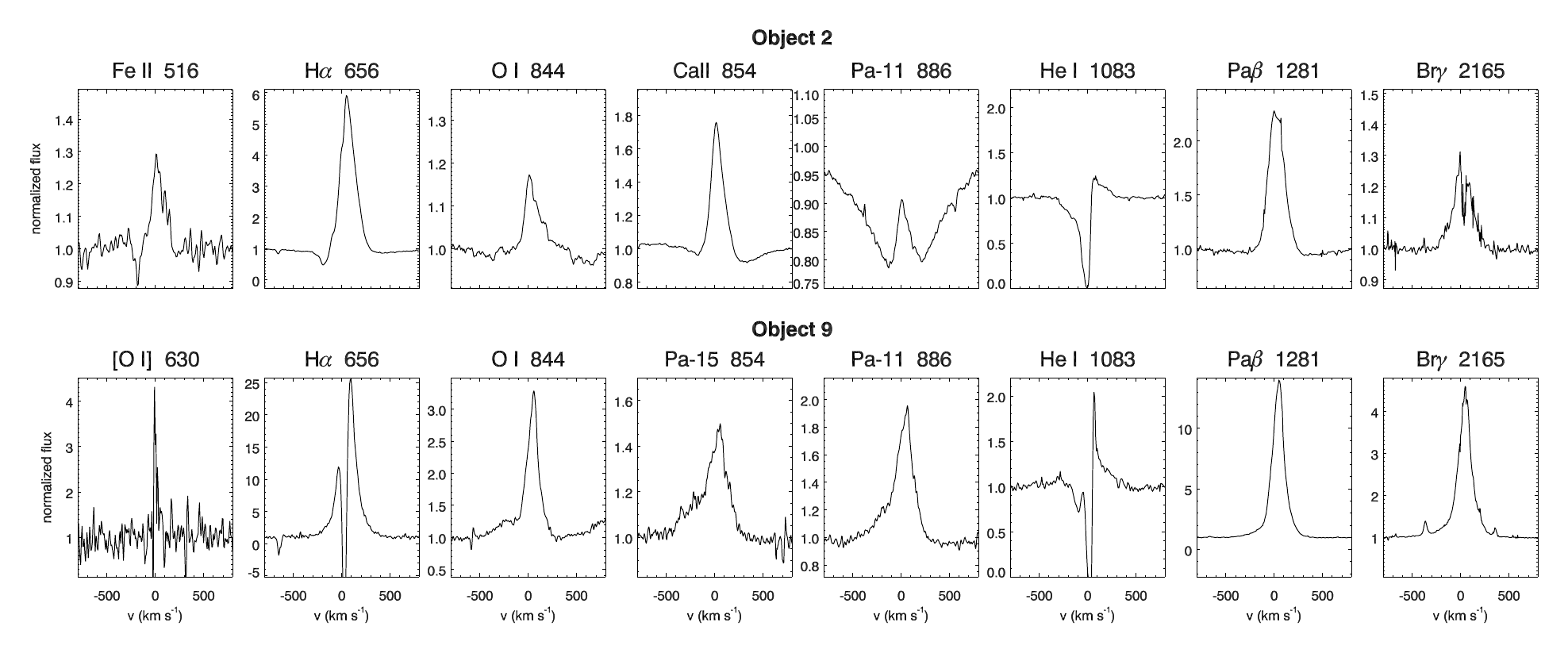}
\caption{Selected emission lines in the spectra of objects 2 (Herbig Ae/PMS star) and 9 (YSO). Wavelengths are in nm. Note the P-Cygni profiles of various lines in  object 2, and the strong blue wings of the H~{\sc i} lines in object 9. The Ca~{\sc ii} and Pa-11 emission lines in object 2 are superposed on photospheric absorption lines. \label{fig:etspectrair}}
\end{figure*}

\begin{figure*}
\centering
\includegraphics[width=\textwidth]{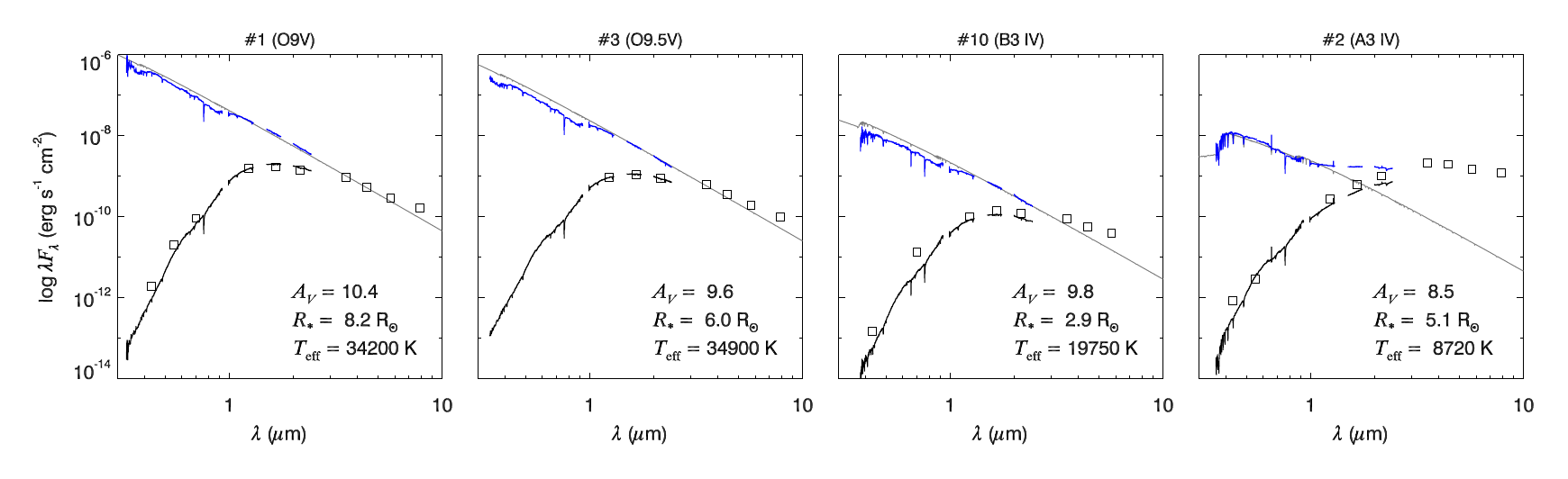} 
\caption{\label{fig:xshseds} SEDs of the four stars that have been classified by their optical spectra. The X-shooter spectrum of each star (black line) is dereddened (blue line) with the $A_V$ value listed in Table~\ref{tab:ostars}. The corresponding Kurucz model is plotted in gray. Optical \citep{Zacharias2005, Monet2003}, SOFI and IRAC photometry is overplotted with squares. The ``kinks'' in the spectra at 0.35 and 0.85~$\mu$m are instrumental features.}
\end{figure*}

\section{Results from spectroscopy}
\label{sec:spectroscopy}

We have obtained SINFONI and X-shooter spectroscopic observations of the brightest objects. In this section we analyze the early-type spectra (X-shooter), the late-type spectra (SINFONI) and the YSOs (X-shooter). Fig.~\ref{fig:maps}e displays the coverage of these observations. The extinction map of the cluster is displayed in Figs.~\ref{fig:maps}b and d; the SINFONI map of the nebular lines is shown in Fig.~\ref{fig:maps}f.

\subsection{Early-type stars: objects 1, 2, 3 and 10}
\label{sec:spectroscopy:et}

The results of the spectral analysis of the four brightest photospheric spectra found in RCW~36 are summarized in Table~\ref{tab:ostars}. In order to constrain the spectral types of the O- and B-type stars (Objects 1, 3 and 10), non-LTE atmosphere models \citep[\texttt{FASTWIND};][]{Puls2005} were fitted to the spectra using a genetic algorithm approach. This allows for the simultaneous determination of the main stellar and wind parameters using a selection of 11 helium and hydrogen lines. We refer to \citet{Mokiem2005} for a description of the algorithm, the parameters and the fitted lines. The resulting best-fit line profiles for both stars are overplotted on the spectra in Fig.~\ref{fig:etspectra}. 

The spectra of objects 1, 3 and 10 have no spectral signatures of circumstellar material, nor anomalously strong mass loss ($\dot{M} < 10^{-7}$~M$_\odot$~yr$^{-1}$) or rotation. We determine optical spectral types of the O stars by quantitative EW measurements following \citet{Conti1971} and \citet{Mathys1988, Mathys1989}, as revised by Sana et al. (in prep.). The spectral types of objects 1 and 3 are O8.5\,--\,9.5~V and O9.5\,--\,B0~V, respectively. Object 10 is a B2\,--\,3.5~IV star based on comparison with synthetic spectra from \citet{Munari2005}. Its surface gravity is not well constrained by the atmosphere fitting and may indeed be lower than the ZAMS value. This could indicate that the star is still in a PMS contraction phase, although it is already on the ZAMS. Note that the narrow, deep central absorption in the H~{\sc i} lines is due to oversubtraction of the nebular spectrum. 

Object 2 is classified by comparing its spectrum in the region $380-700$~nm to synthetic spectra from \citet{Munari2005} to obtain a spectral classification. Using the Ca~K line as the main temperature indicator in the optical, a spectral type A2\,--\,A4 is determined. Based on the shape of the Balmer line wings, the luminosity class is IV. Object 2 exhibits a significant excess emission starting at 1~$\mu$m and a flat (class II disk) SED.  For its placement in the HRD, the absolute magnitude $M_K$ is corrected for the intrinsic infrared excess (amounting to 1.21 mag in $K$). Some of the emission lines in its spectrum (H~{\sc i}, He~{\sc i}, Fe~{\sc ii}) have asymmetric or P-Cygni profiles, indicating the presence of a stellar wind (See Fig.~\ref{fig:etspectrair}). The emission profiles of higher H~{\sc i} transitions are more symmetric and double-peaked, pointing to their origin in a high-density medium (i.e. an inner gas disk). Based on the above, we classify object 2 as a Herbig Ae star. Its radius is ``bloated'' with respect to its main sequence size, consistent with what is expected \citep{Palla1993} and observed \citep{Ochsendorf2011} in intermediate-mass PMS stars. 

The extinction $A_V$ towards objects 1, 2, 3 and 10 is determined within 0.5~mag uncertainty by fitting the slope of the SED to a Kurucz atmospheric model \citep{Kurucz1993} in the entire X-shooter wavelength range (for object 2 up to 1~$\mu$m, see Fig.~\ref{fig:xshseds}). Finally, the stellar radius is determined by scaling the observed flux to the flux at the stellar surface given by the model; this scaling is degenerate with the distance estimate. With the spectral types and confirmed ZAMS nature of the OB stars, the spectroscopic parallax method is used to estimate the distance towards the cluster. For this we use the $K_{\rm s}$-band magnitudes and extinction of objects 1, 3 and 10, and adopt the absolute magnitude calibration defined by the ``observational'' scale defined in \citet{Martins2006}. The results (Table~\ref{tab:ostars}) are in the range of 0.6\,--\,0.8~kpc, consistent with the distance of 0.7~kpc obtained from literature and with the spectroscopic distance estimates by \citet{Bik2005}. The error in the spectroscopic distance determination is dominated by the error in $A_V$.

A radial velocity of $v_{\rm LSR}=9\pm16$~km~s$^{-1}$ could be determined from the photospheric lines of object 3. This is in concurrence with the nebular velocity $v_{\rm LSR}=7.5$~km~s$^{-1}$, measured by \citet{Bronfman1996} from the CS(2-1) emission line. Based on its colors and magnitude, object 14 (located outside the SINFONI field) is probably also a main-sequence OB star, but due to its high extinction ($A_V\sim 25$~mag) and location outside the cluster center it is either an unrelated background star, or an ejected ``runaway'' star. This would have to be confirmed by spectroscopy.

\subsection{Late-type stars}
\label{sec:specroscopy:lt}

A total of 138 SOFI point sources with magnitudes between $K_{\rm s}=6.9$~mag and $K_{\rm s}=16.5$~mag have an associated point source in the SINFONI continuum images. The observations provide an $H$- and $K$-band spectrum  of every source with a spectral resolution of $R \sim 1500$. We have classified the spectra with a signal-to-noise ratio S/N $> 20$ in the $K$-band (corresponding to $K_{\rm s} < 14$~mag); for these a spectral type could be determined within two subtypes. As a result, 47 sources are assigned a spectral type and luminosity class. See Fig.~\ref{fig:maps}e for their location; part of the central cluster is not covered by the SINFONI observations. Table~\ref{tab:dataoverview} summarizes the spectral types and characteristics of the classified sources; their SINFONI spectra are displayed in Fig.~\ref{fig:sinfonispectra}.

Most of these stars have photometric spectral types later than F. The late-type spectra are compared with reference spectra from \citet{Cushing2005} and \citet{Rayner2009}. The spectral classification is based on atomic and molecular absorption lines (e.g., Mg~{\sc i}, Na~{\sc i}, Ca~{\sc ii}, and CO). The depth of the CO lines at 2.3~$\mu$m serves as an indicator for the luminosity class. The photometric spectral type of object 26 (K4~V) agrees with the spectral type determined from its X-shooter spectrum, which was obtained using the region around the Ca~{\sc ii} triplet at 850~nm. Many spectra exhibit H~{\sc i} emission, particularly in the Br$\gamma$ line; the spectra for which this emission appears as a point source in the line map are indicated in Table~\ref{tab:dataoverview}.

The uncertainty in the spectral classification (typically one subtype) is reflected in the errors on the effective temperature, $T_{\rm eff}$, and on the extinction, $A_V$. The effective temperature is determined by the calibration in \citet{Kenyon1995}. Since this calibration  overestimates $T_{\rm eff}$  \citep[by $\sim 500$~K for G to $\sim 200$~K for mid-K,][]{Cohen1979} for PMS stars, a correction was made which contributes significantly to the error budget. Using the intrinsic colors listed by \citet{Kenyon1995} and the $JHK_{\rm s}$-colors from SOFI, we determine $A_V$ for every classified source. Where available, we based the determination of $A_V$ on the $J$- and $H$-band fluxes so as to avoid contamination of continuum excess emission caused by circumstellar material. The thus derived $A_V$ estimates agree reasonably well with the values derived from dereddening to the cTTS locus (Sect.~\ref{sec:photometry:sofi}); their mean ratio is $\langle A_{V,\rm spectra}/A_{V,\rm cTTS} \rangle = 1.2 \pm 0.4$. The spread in these values can in large part be explained by the spread in the intrinsic colors of cTTS stars found by \citet{Meyer1997}.

The above procedure results in an absolute magnitude $M_K$ for every classified star, from which the stellar radius can also be determined. Also, the stellar mass and age were obtained from the absolute magnitudes and effective temperatures, by interpolating between evolutionary tracks and isochrones; see Sect.~\ref{sec:stellarpop:age}. 

\subsection{Young Stellar Objects}
\label{sec:spectroscopy:yso}

In this section, we discuss the spectra of three YSOs in RCW~36 which stand out by their infrared brightness and/or associated outflows: objects 4, 9 and 97. The spectra of 4 and 97 are also discussed in detail in \citet{Ellerbroek2013}; a selection of emission lines from object 9 is displayed in Fig.~\ref{fig:etspectrair}.

The spectrum of object 4 (also known as 08576nr292) exhibits no photospheric features, and is dominated by continuum emission from a circumstellar disk. It contains many emission lines that originate in a disk-jet system. This intermediate-mass YSO has a high accretion rate \citep[$\dot{M}_{\rm acc} \sim 10^{-6}$ M$_\odot$~yr$^{-1}$,][]{Ellerbroek2013}, and is associated with the Herbig-Haro jet \object{HH~1042}, demonstrating its current accretion activity. For an extensive study of object 4 (08576nr292) and its disk-jet system, we refer the reader to \citet{Ellerbroek2011, Ellerbroek2013}. We adopt the $A_V=8.0\pm1.0$~mag based on SED fitting in the former paper. Object 4 is classified as a class 0/I YSO, but it could also be class II as its red bands may be contaminated by emission from the surrounding cloud.

Object 97 (08576nr480) is another class 0/I YSO associated with a jet (\object{HH~1043}), but it is much more embedded than object 4. Objects 4 and 97 are both superposed on the filamentary structures west of the central cluster, demonstrating star formation is ongoing in this region. Their jets contain emission lines from high ionization species like [S~{\sc iii}] and [O~{\sc ii}], which are also detected in the ambient nebular spectrum. This medium is probably ionized by the central O stars \citep{Ellerbroek2013}.

Object 9 is a class 0/I YSO whose severely reddened spectrum exhibits many emission lines (predominantly H~{\sc i}) with asymmetric line profiles with a strong blue-shifted ``wing'' with velocities up to $\sim 500$~\kms indicative of an outflow (Fig.~\ref{fig:etspectrair}). However, the blue wing is only detected in lines associated with high densities and not in forbidden lines, indicating the wind is dense and optically thick, shrouding the stellar photosphere. A dust shell or disk might further obscure the central object, although no CO bandhead emission is seen in this object, unlike in objects 4 and 97. 


\begin{table}[!h]
\begin{minipage}[c]{\textwidth}

\caption{\normalsize{Properties of the H {\sc ii} gas.}}
\renewcommand{\arraystretch}{1.4}
\renewcommand{\footnoterule}{}
\begin{tabular}{lllll}
\hline 
\hline
Region & Offset\footnote{R.A. and Dec. offsets from object 1, in $''$.} & Size & $A_V$ (mag) & He\one~1.70~$\mu$m/Br$\gamma$ \\
(Fig.~\ref{fig:maps}f) &  ($\arcsec$) & ($\arcsec^2$) & (Br10 / Br$\gamma$) & ($A_V$ corr.)\\ 
\hline

I & $3, -44$ & 14 $\times$ 3 & 13.1 $\pm$  0.4 &0.070 $\pm$ 0.007 \\ 
II & $-50, 4$ & 3 $\times$ 2 & 19.6 $\pm$  0.8 &0.059 $\pm$ 0.015 \\ 
III & $-66, 16$ & 1 $\times$ 3 &  $> 37 $ & $ \dots $ \\ 
IV  & $-8, 25$ & 7 $\times$ 14 &  7.3 $\pm$  1.6 & $ < 0.07 $ \\ 
V  & $-14, -7$ & 10 $\times$ 4 & \dots & $\dots $ \\

\hline
\vspace{-20pt}
\end{tabular} 
\label{tab:nebular}

\end{minipage}
\end{table}

\subsection{Extinction and nebular spectrum}
\label{sec:spectroscopy:nebular}

An extinction map (Fig.~\ref{fig:maps}b,\,d) was produced based on the $JHK_{\rm s}$ colors of the SOFI detections. The method for constructing the extinction map was based on that described by \citet{Lombardi2001}, which is in turn based on the color excess method first presented by \citet{Lada1994}. We consider sources for which $JHK_{\rm s}$-photometry exists and exclude foreground sources (Sect.~\ref{sec:photometry:sofi}) and sources for which an intrinsic near-infrared excess is detected (Sect.~\ref{sec:photometry:irac}). For every source, $A_V$ is determined by dereddening to the ZAMS (see Fig.~\ref{fig:soficolors}a). These values are represented by the colored dots in Fig.~\ref{fig:maps}b. Then a spatial grid is defined with a spatial resolution of 4$'' \times 4''$. For every grid point, an $A_V$ value was calculated by taking the weighted average of $A_V$ of the 20 nearest neighboring stars. The weight used is the inverse squared distance to the nearest neighbors, with a minimum of $8''$ (a smoothing parameter).

The spatial resolution is variable across the extinction map, depending on the local stellar surface density. The effective resolution element (defined as the mean distance to the 20 nearest neighbors) decreases radially from $\sim6''$ in the central arcminute of Fig.~\ref{fig:maps}b to $\sim 25''$ at the edges of the map. The error in the extinction measurement of a gridpoint can be expressed as the square root of the variance in $A_V$ of the 20 nearest neighbors, weighted with their inverse squared distance. We find that the relative error in $A_V$ varies between $20-30\%$ across the map, increasing up to $60\%$ in areas $\sim3'$ west and southeast of the center, where not many sources are found. However, the uncertainty is probably dominated by the systematic error in the adopted extinction law ($R_V$ may be higher, and vary across the cluster) and the fact that some sources may possess an intrinsic infrared excess. Both cases would result in an overestimate of $A_V$.

A gradient is visible from high extinction ($A_V \gtrsim 18$~mag) in the eastern part to lower extinction ($A_V \lesssim 12$~mag) westward of the filamentary structures. As this western part is also where the 500~$\mu$m flux peaks (see \S \ref{sec:stellarpop}), the apparent low extinction value is due to the fact that we observe only those sources which are at the forefront of the molecular cloud. In this region, the SOFI sample is thus biased toward sources which are on average less embedded than the rest of the cluster population. Therefore low-extinction regions in Fig.~\ref{fig:maps}b may in fact have a large column density, as the completeness of the dataset is affected by the presence of cold dust. Also, isolated sources that have a very different extinction compared to their nearest neighbors cause a local maximum or minimum in the extinction map, which should be attributed to circumstellar rather than interstellar material. No trend is found between the extinction and the evolutionary status of the sources.

A more reliable estimate of the extinction of individual sources can be obtained by spectral classification, as in that case the intrinsic colors are known. The 47 sources for which this is the case are overplotted on the extinction map in Fig.~\ref{fig:maps}d. Most of the extinction values derived by spectroscopy are similar to the local value of the extinction map. A few sources have a lower individual $A_V$ than the ambient value. This may be because the sources of which a photospheric spectrum could be classified are typically located on the forefront of the cluster. Alternatively, the ambient extinction may be overestimated for the reasons mentioned above.

\begin{figure}[!h]
\centering
\includegraphics[width=0.7\columnwidth]{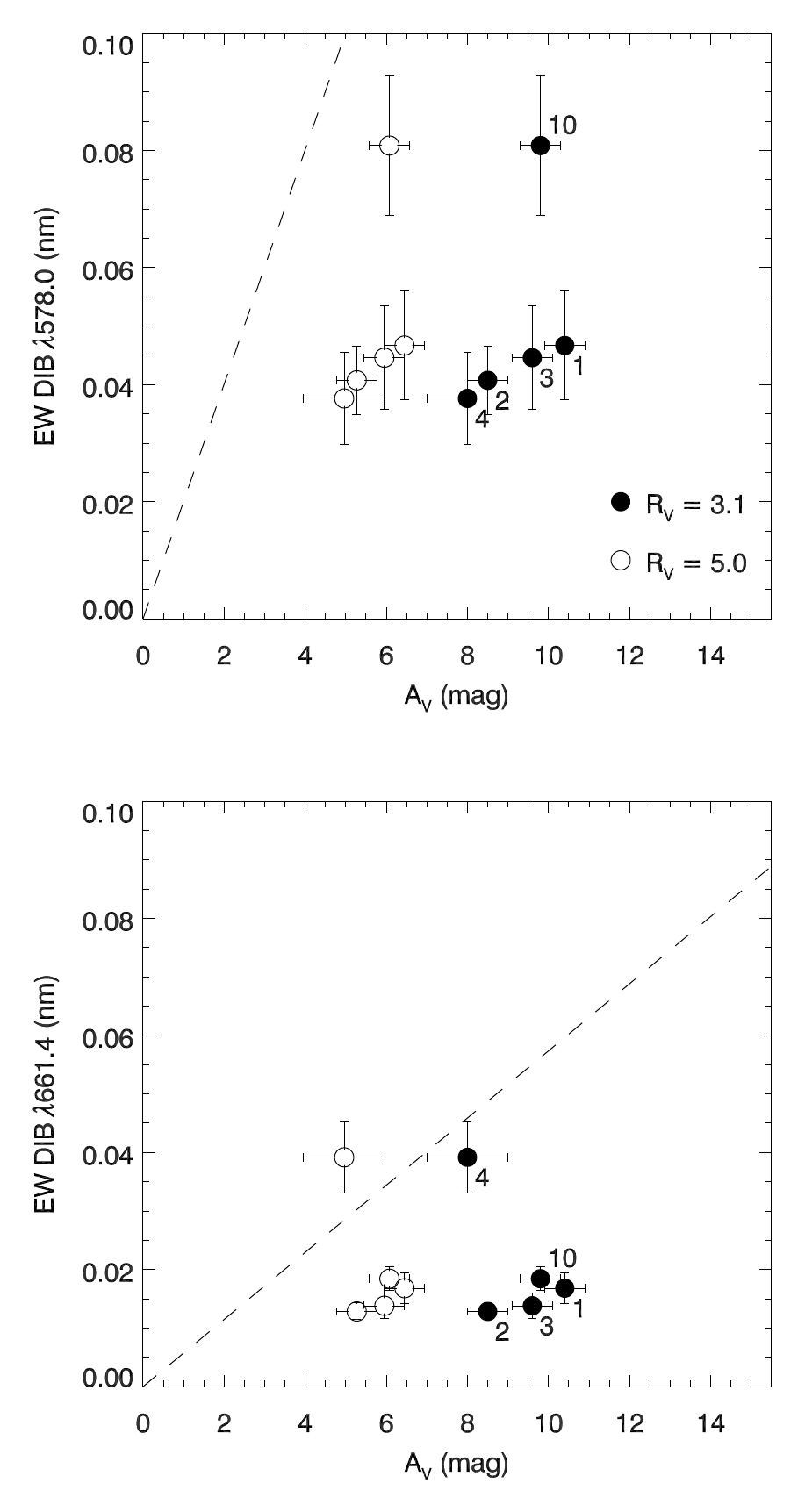}
\caption{Equivalent width of the two strongest diffuse interstellar bands versus extinction for five objects. The dashed line indicates the linear regression in the sample of \citet{Vos2011}. The apparently weak DIB strength may reflect a high value of $R_V$. \label{fig:dibs}}
\end{figure}

An independent estimate of $A_V$ may be obtained by using the observed correlation between the equivalent width (EW) of diffuse interstellar bands (DIBs) and extinction \citep[e.g.][]{Herbig1993, Vos2011}. Fig.~\ref{fig:dibs} shows the correlation between the EW of the two strongest DIBs and $A_V$ (see Sect.~\ref{sec:spectroscopy:et}) of the spectra in which DIBs were detected. The DIB strength to extinction ratios of these spectra are comparable, even though the extinction towards objects 2 and 4 is enhanced by circumstellar material. However, these ratios deviate from the correlation found by \citet{Vos2011}. This discrepancy may reflect an underestimate of $R_V$, which is seen to be higher than the average Galactic value of 3.1 in lines of sight towards star-forming regions \citep[see e.g.][]{Cardelli1989, Hoffmeister2008, Dahlstrom2013}. Alternatively, the DIB carrier(s) may be less abundant in star forming regions, which would provide a constraint on its nature. The limited spread in extinction values in our sample prevent us from testing these possibilities.

The cluster extinction can also be determined by tracers in the nebular spectrum.  Nebular spectra were extracted at five subregions of the H~{\sc ii} region in the SINFONI field of view (Fig.~\ref{fig:maps}f). The locations and sizes of these regions are listed in Table~\ref{tab:nebular}. We have measured the fluxes of selected lines to calculate the extinction and the temperature of the radiation field. In an H~{\sc ii} region, a deviation of the predicted Br10/Br$\gamma$ flux ratio (0.33) is mainly dependent on $A_V$ \citep{Storey1995}. The third column in Table~\ref{tab:nebular} lists the derived extinction values. Regions I--III have an increasing amount of extinction, with the highest extinction at the location where the \textit{Spitzer} mid-infrared flux also peaks (Region III). This is consistent with the high column density at this location \citep{Hill2011}. The region north of the O stars (IV) has a low amount of extinction, while region V has too low S/N to determine the extinction. The $A_V$ values derived by the Br10/Br$\gamma$ ratio (Table~\ref{tab:nebular}) are higher than those derived in the extinction map, possibly because these nebular lines originate in the medium behind the stellar population. The extinction values in regions I, II and IV agree within error with those calculated with the color excess method (Fig.~\ref{fig:maps}b).

The  spectrum of the ionized nebula also provides an estimate on the temperature of its ionizing source. \citet{Lumsden2003} predict the (extinction-corrected) value of the He~{\sc i} 1.70~$\mu$m/Br$\gamma$ line flux ratio as a function of the temperature of the ionizing star. Only in regions I and II the He~{\sc i} 1.70~$\mu$m flux is bright enough; the derived ratios ($0.070 \pm 0.007$ and $0.059 \pm 0.015$) are consistent with a temperature of $34000 \pm 300$~K. This coincides with the temperature determined from the optical spectra of objects 1 and 2 (Sect.~\ref{sec:spectroscopy:et}). This corroborates the finding of \citet{Verma1994} that the emergent flux of the nebula is dominated by the contribution from the central O stars \citep[see also][]{Bik2005}.

\begin{figure*}[!ht]
\centering
\includegraphics[width=0.4\textwidth]{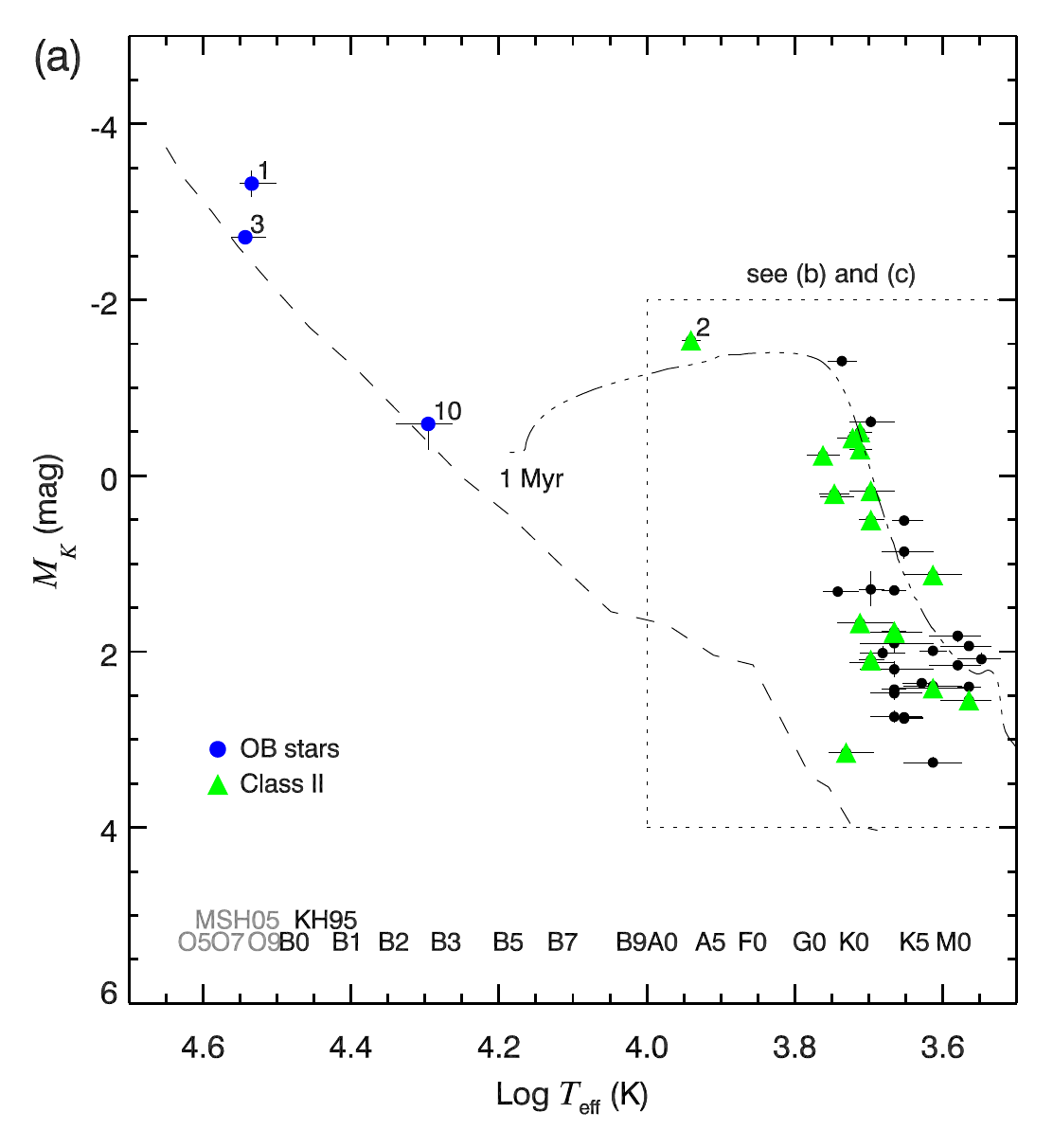} \hspace{0.1cm}
\includegraphics[width=0.41\textwidth]{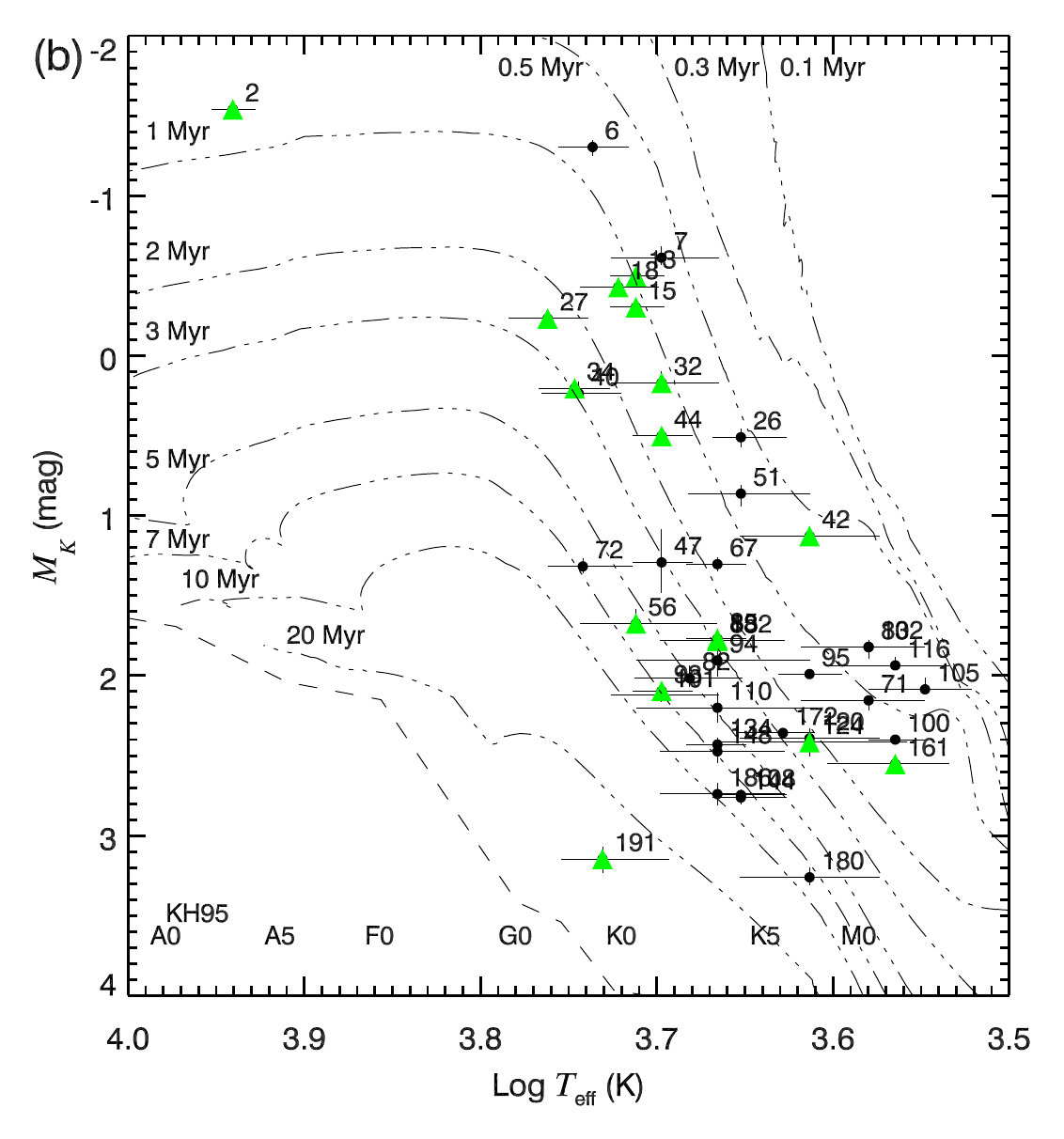}\\
\includegraphics[width=0.41\textwidth]{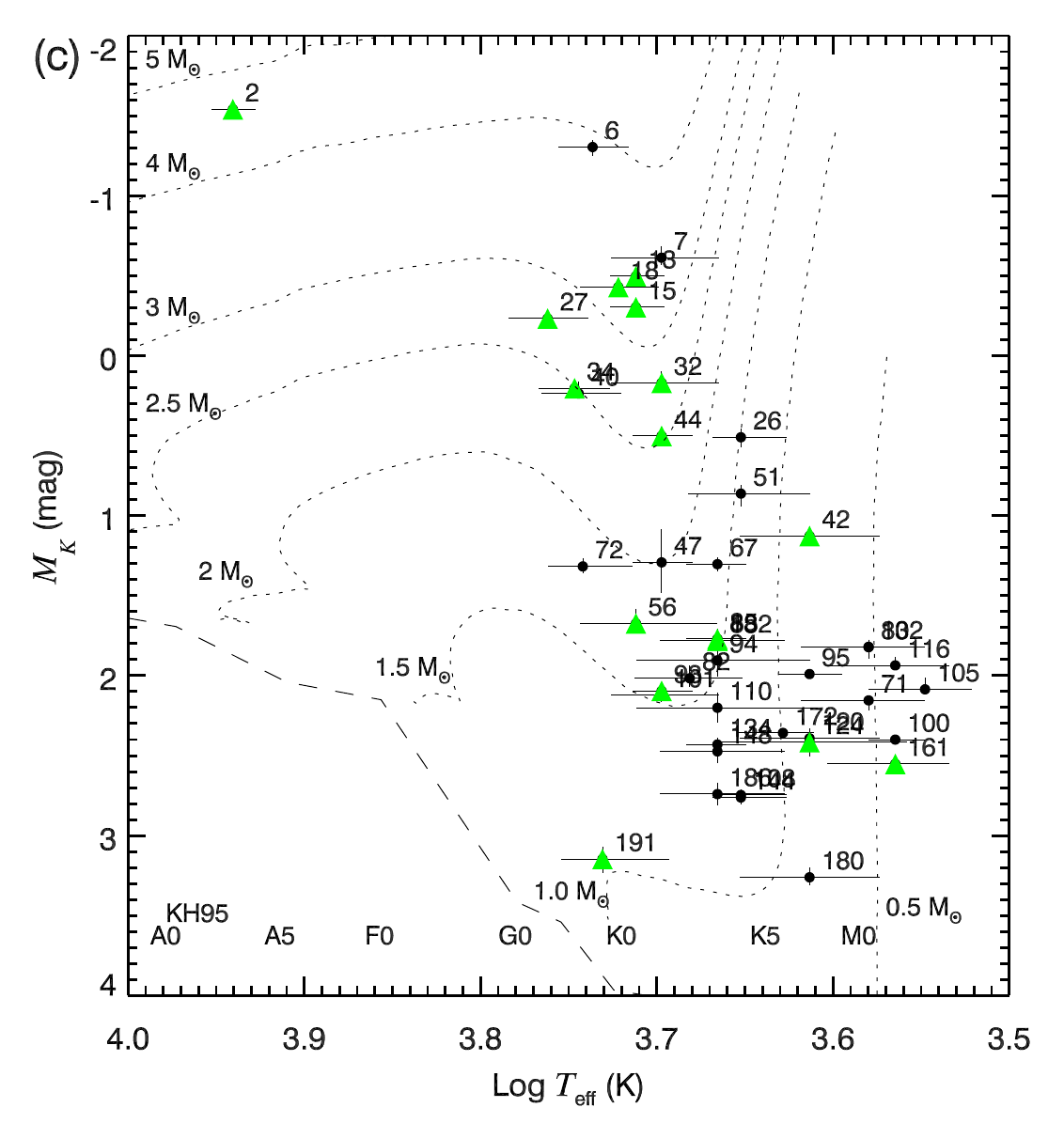} 
\includegraphics[width=0.415\textwidth]{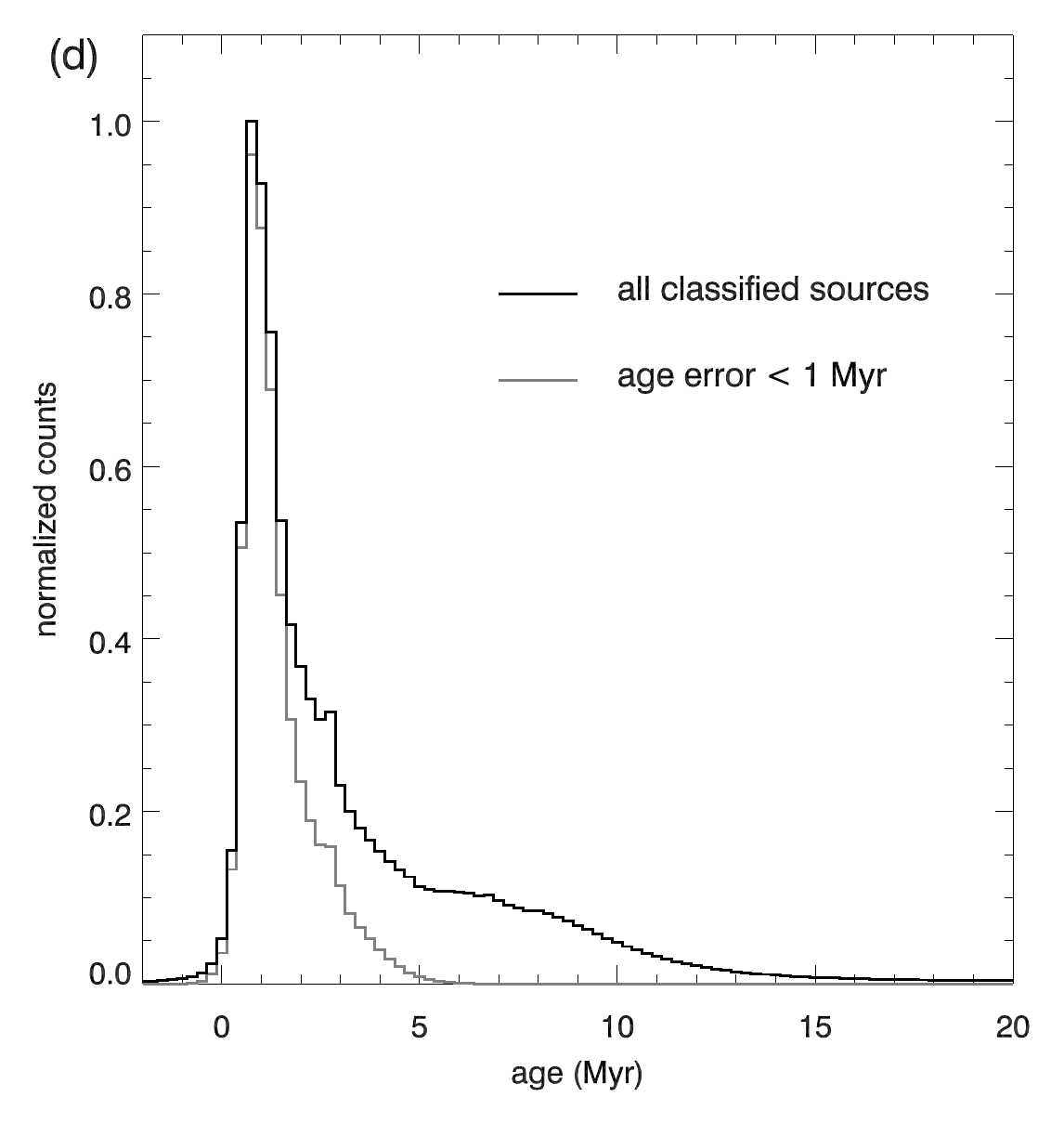}
\caption{\label{fig:hrd} (\textit{a}) Hertzsprung-Russell Diagram including sources in the SINFONI field of view. The $M_K$ value of object 2 is corrected for its intrinsic infrared excess. (\textit{b}) and (\textit{c}) Detail of (\textit{a}), with PMS isochrones (dash-dotted lines) and evolutionary tracks (dotted lines) taken from \citet{DaRio2009}, calculated from the evolutionary models of \citet{Siess2000}. Spectral types are indicated following the observational calibrations by \citet[][O stars]{Martins2006} and \citet[][KH95, B0 and later]{Kenyon1995}. (\textit{d}) Age distribution of the classified stars (see text). The distribution peaks at $1.1 \pm 0.6$~Myr. When only considering sources with errors $<1$~Myr, the ``tail'' towards older ages disappears.}
\end{figure*}

\begin{figure*}[]
\centering
\includegraphics[width=0.98\textwidth]{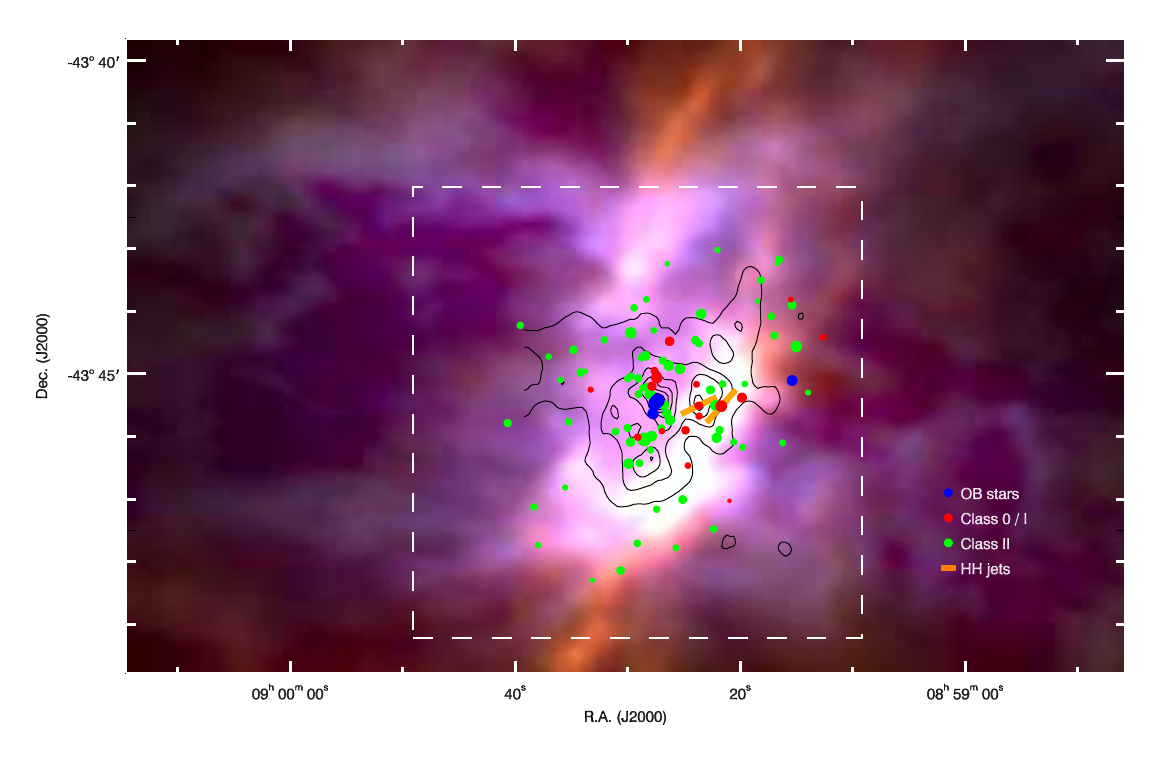}
\caption{\label{fig:mapherschel} Detail of Fig.~\ref{fig:overview} (blue: PACS 70 $\mu$m, green: PACS 160 $\mu$m, red: SPIRE 250 $\mu$m) showing the location of the stellar population with respect to the large-scale bipolar nebula and ring-like structure (MTHM13). Symbol sizes are scaled to the $K_{\rm s}$-band magnitude. Most bright class II objects (green) are located along the major axis of the ring, with the OB stars (blue) in the center. Along the western ridge -- the inner rim of the ring -- some class 0/I objects (red) are found, including those powering two HH jets. Fainter class II sources are located all over the cluster. The stellar surface number density (contours, see text) is highest inside and along the ring and peaks significantly at the location of the OB stars. The contours decrease linearly from 2500~pc$^{-2}$ to 250~pc$^{-2}$. The boxed region indicates the SOFI field of view.}
\end{figure*}

\section{The star formation history of RCW 36}
\label{sec:stellarpop}

In this section, we summarize and interpret the results of the analysis of the stellar population, extinction and nebular properties. We first present the HRD and discuss the age of the stellar population (Sect.~\ref{sec:stellarpop:age}). We then compare our findings with the far-infrared observations of the large-scale molecular cloud in which RCW~36 is embedded and propose a scenario for the star formation history (Sect.~\ref{sec:stellarpop:cloud}).

\subsection{Stellar population: age and spatial distribution}
\label{sec:stellarpop:age}

In the previous section we have derived and discussed the stellar parameters of the sources in RCW~36, summarized in Table~\ref{tab:dataoverview}. With this information we can construct a HRD (Fig.~\ref{fig:hrd}a--c). The stellar population is scattered along the 1 and 2 Myr PMS isochrones, with the O and B stars (objects 1, 3 and 10) already having arrived on the main sequence. Object 10 is possibly a naked and bloated PMS B star based on its apparently low surface gravity, although it is already (nearly) on the main sequence. Object 2 is the brightest class II object that was classified and will likely be the first PMS star arriving on the main sequence after object 10, within the next 0.5~Myr. Along the evolutionary tracks, these stars are followed by the population of low-mass stars, the brightest ones of which are confirmed as class II objects and possess a disk SED similar to object 2.

The PMS age distribution of the classified stars is displayed in a generalized histogram (Fig.~\ref{fig:hrd}d). For every classified star, a gaussian profile centered on its age (derived from the isochrones in Fig.~\ref{fig:hrd}b) with the age error as its width, was added to this histogram. These profiles were normalized so that every star contributes the same area to the histogram. The age distribution peaks at $1.1 \pm 0.6$~Myr, which is the weighted mean age of the stellar population. Many sources have ages within this range, while a number of faint, low-mass ($M<2$~M$_\odot$) sources are located around older isochrones; this is reflected by the ``tail'' towards high ages in Fig.~\ref{fig:hrd}d. However, this ``tail'' disappears when only considering sources with age errors $<1$~Myr (which make up about 75\% of the sample). Thus, the age distribution is consistent with being the result of the large uncertainties in the ages of the fainter sources, which arise because of the dense spacing of isochrones in this region of the HRD. This does not necessarily imply the presence of an older generation of stars.

No significant age gradient can be found in the spatial distribution of classified stars. Based on the position of the bright, centrally located sources in the HRD we infer that the most massive stars in the cluster are in the center of the cluster, as was also concluded by \citet{Baba2004}. 

\subsection{Comparison with molecular cloud structure}
\label{sec:stellarpop:cloud}

RCW~36 is embedded in an elongated dense filamentary structure of $\sim 10$~pc length \citep{Hill2011}. The far-infrared observations with \textit{Herschel} reveal a bipolar nebula extending EW, with a ring-like structure with a major axis of 2~pc at its center. MTHM13 propose that this morphology is the result of a ``blowout'': the ionizing radiation of massive stars cleared out the central region of the ring, ionized its inner edges, and heated the nebula on either side of the ring resulting in the bipolar nebula. Such parsec-scale bipolar molecular outflows are commonly associated with massive star-forming regions \citep{Arce2007}. According to the modeling by MTHM13, the timescale for the currently observed ring and bipolar nebula to form out of an initial filament is less than 1~Myr.

Our study of the stellar population shows that the massive and intermediate-mass stars in the center of RCW~36 have the right age and characteristics for having shaped the ring-like structure and bipolar nebula. The ring is likely blown out by the stellar winds of the O stars. The stellar parameters of objects 1 and 3 yield a combined wind luminosity of 10$^{34}$~erg~s$^{-1}$ \citep{Vink2000}. Assuming an initial ISM density of 100~cm$^{-3}$ and a lifetime of 1~Myr, such a wind would have blown out a cavity with radius $\sim 1$~pc  \citep[][equation 4.4]{Koo1992}, which is the observed size of the ring. Radiation escapes into the regions east and west of the ring, which have a lower density, resulting in the bipolar nebula.

Fig.~\ref{fig:mapherschel} shows the location of the stellar population with respect to the large-scale cloud structure. The stellar surface number density, the number of SOFI detections per pc$^2$, is calculated as the inverse of the squared mean distance to the 20 nearest neighbors on a grid with a spatial resolution of 5$''$. It peaks at the location of the central O stars (at $\sim 2500$~pc$^{-2}$) and decreases outward (down to $\sim 100$~pc$^{-2}$ at the edge of the SOFI field of view). The ring structure (with its major axis extending NS) found by MTHM13 is clearly visible in this image, as well as the bipolar nebula extending EW. The O stars are located along the major axis of the ring, although south of the center. Within the ring, around the O stars, a co-evolving retinue of intermediate-mass (1-5 M$_\odot$) stars is found. Class 0/I objects, including the two jet sources HH~1042 and HH~1043, are found mainly along the ring structure, although some (e.g. object 9) also occupy the central region. Fainter class II objects are spread over the cluster; most are found within or along the ring structure.

Star formation in RCW~36 is currently ongoing. Some protostellar cores \citep[][MTHM13]{Giannini2012} and a UCHII region \citep{Walsh1998} are found in the region where the far-infrared flux peaks (region II/III, Fig.~\ref{fig:maps}f). The column density in this region is very high \citep{Hill2011}, consistent with the fact that radiation only escapes from its surface and not from the embedded sources. It is entirely possible that new massive stars are being formed within this thick column of dust. Star formation is actively taking place at the forefront of this region, as demonstrated by the presence of two HH objects \citep{Ellerbroek2013}. Their positioning with respect to the cloud is reminiscent of  the HH~901 and HH~902 jets emerging from the pillars north of the cluster Trumpler 14 in the Carina Nebula \citep{Smith2010}.

\section{Conclusions}
\label{sec:conclusions}

We have presented a multi-wavelength photometric and spectroscopic study of the massive star-forming region RCW~36. The detailed characterization of the stellar population has resulted in a formation scenario which is in line with (and can partly explain) the current large-scale structure of the molecular cloud. Our most important findings are summarized below:

\begin{itemize}
\item RCW~36 contains a young PMS stellar population; about half of the classified objects have an intrinsic infrared excess. This suggests that many sources in the cluster have disks, which would be consistent with its young age.
\item The radiative energy output is dominated by two O stars (objects 1 and 3) in the cluster center, which illuminate the filamentary edges of the dissipated molecular cloud. These objects, along with a B star (object 10), are located on or close to the ZAMS and are thus probably co-evolving with the lower-mass PMS population.
\item A Herbig Ae star (object 2) is the most massive object on its way to the main sequence; its bloated radius and circumstellar disk confirm its PMS nature.
\item The classified stellar population has a mean PMS age of $1.1\pm0.6$~ Myr. A number of low-mass ($M<2$~M$_\odot$) objects appear to be older, although large uncertainties exist in their age determination. Our findings are consistent with a single-age population; the uncertainties are too large to make quantitative statements about an age spread in this cluster. 
\item A likely scenario for the star formation history of the cluster is obtained by combining the studies of the molecular cloud and the stellar population. The brightest (most massive) objects that formed 1~Myr or more ago, are found along the ``sheet'', with the O stars in the middle of a ring-shaped structure, which may be shaped by the central stars' radiation field. The timescale for the formation of the ring agrees with the age of the central stellar population. 
\item While an embedded forming massive stellar population may exist beyond a large column of dust (the purple circles in Fig.~\ref{fig:maps}a, and the UCHII), a new generation of lower-mass stars is observed at the forefront of this region (objects 4 and 97). Other class 0, I and II YSOs are found in the cluster center as well as the periphery.
\end{itemize}

RCW~36 is revealed as a very rich cluster which can serve as a excellent ``laboratory'' to study sequential star formation. The impact of newly formed massive stars on the ambient molecular cloud is clearly exposed. Our study suggests that the surrounding intermediate-mass stellar population has an age compatible with the dynamical age of the cluster. The presence of an even younger generation of protostellar objects (massive protostellar cores, a UCHII region, YSOs and protostellar jets) shows that RCW~36 is an active site of star formation, which will considerably reshape the molecular cloud within the next millions of years.

\begin{acknowledgements}
The authors thank the anonymous referee for providing useful comments and suggestions that helped improve the paper. A.B. acknowledges the hospitality of the Aspen Center for Physics, which is supported by the National Science Foundation Grant No. PHY-1066293. This research project is financially supported by a grant from the Netherlands Research School for Astronomy (NOVA).
\end{acknowledgements}


\begin{thebibliography}{80}
\expandafter\ifx\csname natexlab\endcsname\relax\def\natexlab#1{#1}\fi

\bibitem[{Abuter {et~al.}(2006)Abuter, Schreiber, Eisenhauer, Ott, Horrobin, \&
  Gillesen}]{Abuter2006}
Abuter, R., Schreiber, J., Eisenhauer, F., {et~al.} 2006, New A. Rev, 50, 398

\bibitem[{{Allen} {et~al.}(2004){Allen}, {Calvet}, {D'Alessio}, {Merin},
  {Hartmann}, {Megeath}, {Gutermuth}, {Muzerolle}, {Pipher}, {Myers}, \&
  {Fazio}}]{Allen2004}
{Allen}, L.~E., {Calvet}, N., {D'Alessio}, P., {et~al.} 2004, \apjs, 154, 363

\bibitem[{{Alvarez} {et~al.}(2004){Alvarez}, {Feldt}, {Henning}, {Puga},
  {Brandner}, \& {Stecklum}}]{Alvarez2004}
{Alvarez}, C., {Feldt}, M., {Henning}, T., {et~al.} 2004, \apjs, 155, 123

\bibitem[{{Arce} {et~al.}(2007){Arce}, {Shepherd}, {Gueth}, {Lee}, {Bachiller},
  {Rosen}, \& {Beuther}}]{Arce2007}
{Arce}, H.~G., {Shepherd}, D., {Gueth}, F., {et~al.} 2007, Protostars and
  Planets V, 245

\bibitem[{{Baba} {et~al.}(2004){Baba}, {Nagata}, {Nagayama}, {Nagashima},
  {Kato}, {Kurita}, {Sato}, {Nakajima}, {Tamura}, {Nakaya}, \&
  {Sugitani}}]{Baba2004}
{Baba}, D., {Nagata}, T., {Nagayama}, T., {et~al.} 2004, \apj, 614, 818

\bibitem[{{Balog} {et~al.}(2007){Balog}, {Muzerolle}, {Rieke}, {Su}, {Young},
  \& {Megeath}}]{Balog07}
{Balog}, Z., {Muzerolle}, J., {Rieke}, G.~H., {et~al.} 2007, ApJ, 660, 1532

\bibitem[{{Bessell} \& {Brett}(1988)}]{Bessell1988}
{Bessell}, M.~S. \& {Brett}, J.~M. 1988, \pasp, 100, 1134

\bibitem[{{Bik}(2004)}]{Bik2004}
{Bik}, A. 2004, PhD thesis, University of Amsterdam

\bibitem[{{Bik} {et~al.}(2005){Bik}, {Kaper}, {Hanson}, \& {Smits}}]{Bik2005}
{Bik}, A., {Kaper}, L., {Hanson}, M.~M., \& {Smits}, M. 2005, \aap, 440, 121

\bibitem[{{Bik} {et~al.}(2006){Bik}, {Kaper}, \& {Waters}}]{Bik2006}
{Bik}, A., {Kaper}, L., \& {Waters}, L.~B.~F.~M. 2006, \aap, 455, 561

\bibitem[{{Bik} {et~al.}(2010){Bik}, {Puga}, {Waters}, {Horrobin}, {Henning},
  {Vasyunina}, {Beuther}, {Linz}, {Kaper}, {van den Ancker}, {Lenorzer},
  {Churchwell}, {Kurtz}, {Kouwenhoven}, {Stolte}, {de Koter}, {Thi},
  {Comer{\'o}n}, \& {Waelkens}}]{Bik2010}
{Bik}, A., {Puga}, E., {Waters}, L.~B.~F.~M., {et~al.} 2010, \apj, 713, 883

\bibitem[{{Bik} \& {Thi}(2004)}]{BikThi2004}
{Bik}, A. \& {Thi}, W.~F. 2004, \aap, 427, L13

\bibitem[{{Blum} {et~al.}(2000){Blum}, {Conti}, \& {Damineli}}]{Blum2000}
{Blum}, R.~D., {Conti}, P.~S., \& {Damineli}, A. 2000, \aj, 119, 1860

\bibitem[{Bonnet {et~al.}(2004)Bonnet, Abuter, Baker, Bornemann, Brown,
  Castillo, Conzelmann, Damster, Davies, Delabre, Donaldson, Dumas, Eisenhardt,
  Elswijk, Fedrigo, Finger, Gemperlein, Genzel, Gilbert, Gillet, Goldbrunner,
  Horrobin, ter Horst, Huber, Hubin, Iserlohe, Kaufer, Kissler-Patig, Kragt,
  Kroes, Lehnert, Lieb, Liske, Lizon, Lutz, Modigliani, Monnet, Nesvadba,
  Patig, Pragt, Reunanen, R{\"o}hrle, Rossi, Schmutzer, Schoenmaker, Schreiber,
  Stroebele, Szeifert, Tacconi-Garman, Tecza, Thatte, Tordo, van~der Werf, \&
  Weisz}]{Bonnet2004}
Bonnet, H., Abuter, R., Baker, A., {et~al.} 2004, The Messenger, 117, 17

\bibitem[{{Bronfman} {et~al.}(1996){Bronfman}, {Nyman}, \&
  {May}}]{Bronfman1996}
{Bronfman}, L., {Nyman}, L.-A., \& {May}, J. 1996, \aaps, 115, 81

\bibitem[{{Cardelli} {et~al.}(1989){Cardelli}, {Clayton}, \&
  {Mathis}}]{Cardelli1989}
{Cardelli}, J.~A., {Clayton}, G.~C., \& {Mathis}, J.~S. 1989, \apj, 345, 245

\bibitem[{{Caswell} \& {Haynes}(1987)}]{Caswell1987}
{Caswell}, J.~L. \& {Haynes}, R.~F. 1987, \aap, 171, 261

\bibitem[{{Cohen} \& {Kuhi}(1979)}]{Cohen1979}
{Cohen}, M. \& {Kuhi}, L.~V. 1979, \apjs, 41, 743

\bibitem[{{Conti} \& {Alschuler}(1971)}]{Conti1971}
{Conti}, P.~S. \& {Alschuler}, W.~R. 1971, \apj, 170, 325

\bibitem[{{Cushing} {et~al.}(2005){Cushing}, {Rayner}, \&
  {Vacca}}]{Cushing2005}
{Cushing}, M.~C., {Rayner}, J.~T., \& {Vacca}, W.~D. 2005, \apj, 623, 1115

\bibitem[{{Da Rio} {et~al.}(2009){Da Rio}, {Gouliermis}, \&
  {Henning}}]{DaRio2009}
{Da Rio}, N., {Gouliermis}, D.~A., \& {Henning}, T. 2009, \apj, 696, 528

\bibitem[{{Dahlstrom} {et~al.}(2013){Dahlstrom}, {York}, {Welty}, {Oka},
  {Hobbs}, {Johnson}, {Friedman}, {Jiang}, {Rachford}, {Sherman}, {Snow}, \&
  {Sonnentrucker}}]{Dahlstrom2013}
{Dahlstrom}, J., {York}, D.~G., {Welty}, D.~E., {et~al.} 2013, \apj, 773, 41

\bibitem[{{Davies} {et~al.}(2011){Davies}, {Hoare}, {Lumsden}, {Hosokawa},
  {Oudmaijer}, {Urquhart}, {Mottram}, \& {Stead}}]{Davies2011}
{Davies}, B., {Hoare}, M.~G., {Lumsden}, S.~L., {et~al.} 2011, \mnras, 416, 972

\bibitem[{Eisenhauer {et~al.}(2003)Eisenhauer, Abuter, Bickert,
  Biancat-Marchet, Bonnet, Brynnel, Conzelmann, Delabre, Donaldson, Farinato,
  Fedrigo, Genzel, Hubin, Iserlohe, Kasper, Kissler-Patig, Monnet, Roehrle,
  Schreiber, Stroebele, Tecza, Thatte, \& Weisz}]{Eisenhauer2003}
Eisenhauer, F., Abuter, R., Bickert, K., {et~al.} 2003, SPIE, 4841, 1548

\bibitem[{{Ekstr{\"o}m} {et~al.}(2012){Ekstr{\"o}m}, {Georgy}, {Eggenberger},
  {Meynet}, {Mowlavi}, {Wyttenbach}, {Granada}, {Decressin}, {Hirschi},
  {Frischknecht}, {Charbonnel}, \& {Maeder}}]{Ekstrom2012}
{Ekstr{\"o}m}, S., {Georgy}, C., {Eggenberger}, P., {et~al.} 2012, \aap, 537,
  A146

\bibitem[{{Ellerbroek} {et~al.}(2011){Ellerbroek}, {Kaper}, {Bik}, {de Koter},
  {Horrobin}, {Puga}, {Sana}, \& {Waters}}]{Ellerbroek2011}
{Ellerbroek}, L.~E., {Kaper}, L., {Bik}, A., {et~al.} 2011, \apjl, 732, L9

\bibitem[{{Ellerbroek} {et~al.}(2013){Ellerbroek}, {Podio}, {Kaper}, {Sana},
  {Huppenkothen}, {de Koter}, \& {Monaco}}]{Ellerbroek2013}
{Ellerbroek}, L.~E., {Podio}, L., {Kaper}, L., {et~al.} 2013, \aap, 551, A5

\bibitem[{Fazio {et~al.}(2004)Fazio, Hora, Allen, Ashby, Barmby, Deutsch,
  Huang, Kleiner, Marengo, Megeath, Melnick, Pahre, Patten, Polizotti, Smith,
  Taylor, Wang, Willner, Hoffmann, Pipher, Forrest, McMurty, McCreight,
  McKelvey, McMurray, Koch, Moseley, Arendt, Mentzell, Marx, Losch, Mayman,
  Eichhorn, Krebs, Jhabvala, Gezari, Fixsen, Flores, Shakoorzadeh, Jungo,
  Hakun, Workman, Karpati, Kichak, Whitley, Mann, Tollestrup, Eisenhardt,
  Stern, Gorjian, Bhattacharya, Carey, Nelson, Glaccum, Lacy, Lowrance, Laine,
  Reach, Stauffer, Surace, Wilson, Wright, Hoffman, Domingo, \&
  Cohen}]{Fazio04}
Fazio, G.~G., Hora, J.~L., Allen, L.~E., {et~al.} 2004, APJS, 154, 10

\bibitem[{{Feigelson} \& {Townsley}(2008)}]{Feigelson2008}
{Feigelson}, E.~D. \& {Townsley}, L.~K. 2008, \apj, 673, 354

\bibitem[{{Giannini} {et~al.}(2012){Giannini}, {Elia}, {Lorenzetti},
  {Molinari}, {Motte}, {Schisano}, {Pezzuto}, {Pestalozzi}, {di Giorgio},
  {Andr{\'e}}, {Hill}, {Benedettini}, {Bontemps}, {di Francesco}, {Fallscheer},
  {Hennemann}, {Kirk}, {Minier}, {Nguyn Lu'O'Ng}, {Polychroni}, {Rygl},
  {Saraceno}, {Schneider}, {Spinoglio}, {Testi}, {Ward-Thompson}, \&
  {White}}]{Giannini2012}
{Giannini}, T., {Elia}, D., {Lorenzetti}, D., {et~al.} 2012, \aap, 539, A156

\bibitem[{{Gutermuth} {et~al.}(2009){Gutermuth}, {Megeath}, {Myers}, {Allen},
  {Pipher}, \& {Fazio}}]{Gutermuth2009}
{Gutermuth}, R.~A., {Megeath}, S.~T., {Myers}, P.~C., {et~al.} 2009, \apjs,
  184, 18

\bibitem[{{Hanson} {et~al.}(1997){Hanson}, {Howarth}, \& {Conti}}]{Hanson1997}
{Hanson}, M.~M., {Howarth}, I.~D., \& {Conti}, P.~S. 1997, \apj, 489, 698

\bibitem[{{Herbig}(1993)}]{Herbig1993}
{Herbig}, G.~H. 1993, \apj, 407, 142

\bibitem[{{Hern{\'a}ndez} {et~al.}(2008){Hern{\'a}ndez}, {Hartmann}, {Calvet},
  {Jeffries}, {Gutermuth}, {Muzerolle}, \& {Stauffer}}]{Hernandez2008}
{Hern{\'a}ndez}, J., {Hartmann}, L., {Calvet}, N., {et~al.} 2008, \apj, 686,
  1195

\bibitem[{{Hill} {et~al.}(2011){Hill}, {Motte}, {Didelon}, {Bontemps},
  {Minier}, {Hennemann}, {Schneider}, {Andr{\'e}}, {Men'shchikov}, {Anderson},
  {Arzoumanian}, {Bernard}, {di Francesco}, {Elia}, {Giannini}, {Griffin},
  {K{\"o}nyves}, {Kirk}, {Marston}, {Martin}, {Molinari}, {Nguyen Luong},
  {Peretto}, {Pezzuto}, {Roussel}, {Sauvage}, {Sousbie}, {Testi},
  {Ward-Thompson}, {White}, {Wilson}, \& {Zavagno}}]{Hill2011}
{Hill}, T., {Motte}, F., {Didelon}, P., {et~al.} 2011, \aap, 533, A94

\bibitem[{{Hoffmeister} {et~al.}(2008){Hoffmeister}, {Chini}, {Scheyda},
  {Schulze}, {Watermann}, {N{\"u}rnberger}, \& {Vogt}}]{Hoffmeister2008}
{Hoffmeister}, V.~H., {Chini}, R., {Scheyda}, C.~M., {et~al.} 2008, \apj, 686,
  310

\bibitem[{{Hunt-Cunningham} {et~al.}(2002){Hunt-Cunningham}, {Whiteoak}, \&
  {Priestley}}]{Hunt-Cunningham2002}
{Hunt-Cunningham}, M.~R., {Whiteoak}, J.~B., \& {Priestley}, P. 2002, in 8th
  Asian-Pacific Regional Meeting, Volume II, ed. S.~{Ikeuchi}, J.~{Hearnshaw},
  \& T.~{Hanawa}, 145--146

\bibitem[{{Kenyon} \& {Hartmann}(1995)}]{Kenyon1995}
{Kenyon}, S.~J. \& {Hartmann}, L. 1995, \apjs, 101, 117

\bibitem[{{Koo} \& {McKee}(1992)}]{Koo1992}
{Koo}, B.-C. \& {McKee}, C.~F. 1992, \apj, 388, 93

\bibitem[{{Kurucz}(1993)}]{Kurucz1993}
{Kurucz}, R.~L. 1993, VizieR Online Data Catalog, 6039, 0

\bibitem[{{Lada}(1987)}]{Lada1987}
{Lada}, C.~J. 1987, in IAU Symposium, Vol. 115, Star Forming Regions, ed.
  {M.~Peimbert \& J.~Jugaku}, 1--17

\bibitem[{{Lada} \& {Lada}(2003)}]{Lada2003}
{Lada}, C.~J. \& {Lada}, E.~A. 2003, \araa, 41, 57

\bibitem[{{Lada} {et~al.}(1994){Lada}, {Lada}, {Clemens}, \&
  {Bally}}]{Lada1994}
{Lada}, C.~J., {Lada}, E.~A., {Clemens}, D.~P., \& {Bally}, J. 1994, \apj, 429,
  694

\bibitem[{{Liseau} {et~al.}(1992){Liseau}, {Lorenzetti}, {Nisini}, {Spinoglio},
  \& {Moneti}}]{Liseau1992}
{Liseau}, R., {Lorenzetti}, D., {Nisini}, B., {Spinoglio}, L., \& {Moneti}, A.
  1992, \aap, 265, 577

\bibitem[{{Lombardi} \& {Alves}(2001)}]{Lombardi2001}
{Lombardi}, M. \& {Alves}, J. 2001, \aap, 377, 1023

\bibitem[{{Lumsden} {et~al.}(2003){Lumsden}, {Puxley}, {Hoare}, {Moore}, \&
  {Ridge}}]{Lumsden2003}
{Lumsden}, S.~L., {Puxley}, P.~J., {Hoare}, M.~G., {Moore}, T.~J.~T., \&
  {Ridge}, N.~A. 2003, \mnras, 340, 799

\bibitem[{{Maaskant} {et~al.}(2011){Maaskant}, {Bik}, {Waters}, {Kaper},
  {Henning}, {Puga}, {Horrobin}, \& {Kainulainen}}]{Maaskant2011}
{Maaskant}, K.~M., {Bik}, A., {Waters}, L.~B.~F.~M., {et~al.} 2011, \aap, 531,
  A27

\bibitem[{{Martins} \& {Plez}(2006)}]{Martins2006}
{Martins}, F. \& {Plez}, B. 2006, \aap, 457, 637

\bibitem[{{Mathys}(1988)}]{Mathys1988}
{Mathys}, G. 1988, \aaps, 76, 427

\bibitem[{{Mathys}(1989)}]{Mathys1989}
{Mathys}, G. 1989, \aaps, 81, 237

\bibitem[{{Megeath} {et~al.}(2004){Megeath}, {Allen}, {Gutermuth}, {Pipher},
  {Myers}, {Calvet}, {Hartmann}, {Muzerolle}, \& {Fazio}}]{Megeath2004}
{Megeath}, S.~T., {Allen}, L.~E., {Gutermuth}, R.~A., {et~al.} 2004, \apjs,
  154, 367

\bibitem[{{Meyer} {et~al.}(1997){Meyer}, {Calvet}, \&
  {Hillenbrand}}]{Meyer1997}
{Meyer}, M.~R., {Calvet}, N., \& {Hillenbrand}, L.~A. 1997, \aj, 114, 288

\bibitem[{{Minier} {et~al.}(2013){Minier}, {Tremblin}, {Hill}, {Motte},
  {Andr{\'e}}, {Lo}, {Schneider}, {Audit}, {White}, {Hennemann}, {Cunningham},
  {Deharveng}, {Didelon}, {Di Francesco}, {Elia}, {Giannini}, {Nguyen Luong},
  {Pezzuto}, {Rygl}, {Spinoglio}, {Ward-Thompson}, \& {Zavagno}}]{Minier2013}
{Minier}, V., {Tremblin}, P., {Hill}, T., {et~al.} 2013, \aap, 550, A50
  (MTHM13)

\bibitem[{Modigliani {et~al.}(2010)Modigliani, Goldoni, Royer, Haigron,
  Guglielmi, Fran{\c c}ois, Horrobin, Bristow, Vernet, Moehler, Kerber,
  Ballester, Mason, \& Christensen}]{Modigliani2010}
Modigliani, A., Goldoni, P., Royer, F., {et~al.} 2010, Observatory Operations:
  Strategies, 7737, 56, (c) 2010: American Institute of Physics

\bibitem[{{Mokiem} {et~al.}(2005){Mokiem}, {de Koter}, {Puls}, {Herrero},
  {Najarro}, \& {Villamariz}}]{Mokiem2005}
{Mokiem}, M.~R., {de Koter}, A., {Puls}, J., {et~al.} 2005, \aap, 441, 711

\bibitem[{{Monet} {et~al.}(2003){Monet}, {Levine}, {Canzian}, {Ables}, {Bird},
  {Dahn}, {Guetter}, {Harris}, {Henden}, {Leggett}, {Levison}, {Luginbuhl},
  {Martini}, {Monet}, {Munn}, {Pier}, {Rhodes}, {Riepe}, {Sell}, {Stone},
  {Vrba}, {Walker}, {Westerhout}, {Brucato}, {Reid}, {Schoening}, {Hartley},
  {Read}, \& {Tritton}}]{Monet2003}
{Monet}, D.~G., {Levine}, S.~E., {Canzian}, B., {et~al.} 2003, \aj, 125, 984

\bibitem[{{Moorwood} {et~al.}(1998){Moorwood}, {Cuby}, \&
  {Lidman}}]{Moorwood98}
{Moorwood}, A., {Cuby}, J.-G., \& {Lidman}, C. 1998, The Messenger, 91, 9

\bibitem[{{Munari} {et~al.}(2005){Munari}, {Sordo}, {Castelli}, \&
  {Zwitter}}]{Munari2005}
{Munari}, U., {Sordo}, R., {Castelli}, F., \& {Zwitter}, T. 2005, \aap, 442,
  1127

\bibitem[{{Ochsendorf} {et~al.}(2011){Ochsendorf}, {Ellerbroek}, {Chini},
  {Hartoog}, {Hoffmeister}, {Waters}, \& {Kaper}}]{Ochsendorf2011}
{Ochsendorf}, B.~B., {Ellerbroek}, L.~E., {Chini}, R., {et~al.} 2011, \aap,
  536, L1

\bibitem[{{Palla} \& {Stahler}(1993)}]{Palla1993}
{Palla}, F. \& {Stahler}, S.~W. 1993, \apj, 418, 414

\bibitem[{{Preibisch}(2012)}]{Preibisch2012}
{Preibisch}, T. 2012, Research in Astronomy and Astrophysics, 12, 1

\bibitem[{{Puls} {et~al.}(2005){Puls}, {Urbaneja}, {Venero}, {Repolust},
  {Springmann}, {Jokuthy}, \& {Mokiem}}]{Puls2005}
{Puls}, J., {Urbaneja}, M.~A., {Venero}, R., {et~al.} 2005, \aap, 435, 669

\bibitem[{{Rayner} {et~al.}(2009){Rayner}, {Cushing}, \& {Vacca}}]{Rayner2009}
{Rayner}, J.~T., {Cushing}, M.~C., \& {Vacca}, W.~D. 2009, \apjs, 185, 289

\bibitem[{Schreiber {et~al.}(2004)Schreiber, Thatte, Eisenhauer, Tecza, Abuter,
  \& Horrobin}]{Schreiber2004}
Schreiber, J., Thatte, N., Eisenhauer, F., {et~al.} 2004, ASP Conf. Ser. 314,
  Astronomical Data Analysis Software and Systems XIII (San Francisco: ASP),
  380

\bibitem[{{Siess} {et~al.}(2000){Siess}, {Dufour}, \& {Forestini}}]{Siess2000}
{Siess}, L., {Dufour}, E., \& {Forestini}, M. 2000, \aap, 358, 593

\bibitem[{Skrutskie {et~al.}(2006)Skrutskie, Cutri, Stiening, Weinberg,
  Schneider, Carpenter, Beichman, Capps, Chester, Elias, Huchra, Liebert,
  Lonsdale, Monet, Price, Seitzer, Jarrett, Kirkpatrick, Gizis, Howard, Evans,
  Fowler, Fullmer, Hurt, Light, Kopan, Marsh, McCallon, Tam, van Dyk, \&
  Wheelock}]{Skrutskie06}
Skrutskie, M.~F., Cutri, R.~M., Stiening, R., {et~al.} 2006, AJ, 131, 1163

\bibitem[{{Smith} {et~al.}(2010){Smith}, {Bally}, \& {Walborn}}]{Smith2010}
{Smith}, N., {Bally}, J., \& {Walborn}, N.~R. 2010, \mnras, 405, 1153

\bibitem[{Stetson(1987)}]{Stetson87}
Stetson, P.~B. 1987, PASP, 99, 191

\bibitem[{{Storey} \& {Hummer}(1995)}]{Storey1995}
{Storey}, P.~J. \& {Hummer}, D.~G. 1995, \mnras, 272, 41

\bibitem[{{Verma} {et~al.}(1994){Verma}, {Bisht}, {Ghosh}, {Iyengar},
  {Rengarajan}, \& {Tandon}}]{Verma1994}
{Verma}, R.~P., {Bisht}, R.~S., {Ghosh}, S.~K., {et~al.} 1994, \aap, 284, 936

\bibitem[{{Vink} {et~al.}(2000){Vink}, {de Koter}, \& {Lamers}}]{Vink2000}
{Vink}, J.~S., {de Koter}, A., \& {Lamers}, H.~J.~G.~L.~M. 2000, \aap, 362, 295

\bibitem[{{Vos} {et~al.}(2011){Vos}, {Cox}, {Kaper}, {Spaans}, \&
  {Ehrenfreund}}]{Vos2011}
{Vos}, D.~A.~I., {Cox}, N.~L.~J., {Kaper}, L., {Spaans}, M., \& {Ehrenfreund},
  P. 2011, \aap, 533, A129

\bibitem[{{Walborn} \& {Blades}(1997)}]{Walborn1997}
{Walborn}, N.~R. \& {Blades}, J.~C. 1997, \apjs, 112, 457

\bibitem[{{Walsh} {et~al.}(1998){Walsh}, {Burton}, {Hyland}, \&
  {Robinson}}]{Walsh1998}
{Walsh}, A.~J., {Burton}, M.~G., {Hyland}, A.~R., \& {Robinson}, G. 1998,
  \mnras, 301, 640

\bibitem[{{Wang} {et~al.}(2011){Wang}, {Beuther}, {Bik}, {Vasyunina}, {Jiang},
  {Puga}, {Linz}, {Rod{\'o}n}, {Henning}, \& {Tamura}}]{Wang2011}
{Wang}, Y., {Beuther}, H., {Bik}, A., {et~al.} 2011, \aap, 527, A32

\bibitem[{{Watson} \& {Hanson}(1997)}]{Watson1997}
{Watson}, A.~M. \& {Hanson}, M.~M. 1997, \apjl, 490, L165

\bibitem[{{Yamaguchi} {et~al.}(1999){Yamaguchi}, {Mizuno}, {Saito},
  {Matsunaga}, {Mizuno}, {Ogawa}, \& {Fukui}}]{Yamaguchi1999}
{Yamaguchi}, N., {Mizuno}, N., {Saito}, H., {et~al.} 1999, \pasj, 51, 775

\bibitem[{{Zacharias} {et~al.}(2005){Zacharias}, {Monet}, {Levine}, {Urban},
  {Gaume}, \& {Wycoff}}]{Zacharias2005}
{Zacharias}, N., {Monet}, D.~G., {Levine}, S.~E., {et~al.} 2005, VizieR Online
  Data Catalog, 1297, 0

\bibitem[{{Zavagno} {et~al.}(2006){Zavagno}, {Deharveng}, {Comer{\'o}n},
  {Brand}, {Massi}, {Caplan}, \& {Russeil}}]{Zavagno2006}
{Zavagno}, A., {Deharveng}, L., {Comer{\'o}n}, F., {et~al.} 2006, \aap, 446,
  171

\bibitem[{{Zavagno} {et~al.}(2007){Zavagno}, {Pomar{\`e}s}, {Deharveng},
  {Hosokawa}, {Russeil}, \& {Caplan}}]{Zavagno2007}
{Zavagno}, A., {Pomar{\`e}s}, M., {Deharveng}, L., {et~al.} 2007, \aap, 472,
  835

\end{thebibliography}

\appendix
\section{Classified sources in RCW~36}

Table~\ref{tab:dataoverview} contains an overview of the stellar properties of a selected sample of sources which are assigned a photometric spectral type and/or are individually described in the text. Fig.~\ref{fig:sinfonispectra} displays the normalized SINFONI spectra of the same sources. The locations of some of the absorption (and emission) lines used for spectral classification are indicated with vertical lines.

\onecolumn
\scriptsize{
\renewcommand{\arraystretch}{1.4}
\hspace{0cm}\begin{longtable}{lllllllllllll}
\caption{\normalsize{Properties of selected sources in RCW~36.}}\\

\hline
\hline
Object &\multicolumn{2}{c}{\textit{(J2000)} }  & \multicolumn{1}{c}{ } & \multicolumn{3}{c}{\textit{Spectrum} } & \multicolumn{6}{c}{\textit{Properties}}\\
\#  (B05$^a$) & RA (h m s) &   Dec ($^\circ$ $\arcmin$ $\arcsec$) & Data$^b$ & Sp. Type$^c$ & Lada Class$^d$  &  Br$\gamma$ em. & T$_{\rm eff}$ (K)$^e$  & $K_{\rm s}$ (mag) & $M_K$$^f$ (mag) & $A_V$$^g$ (mag) & $R_*$ (R$_{\odot}$)$^h$ & $M_*$ (M$_\odot$)$^i$ \\
\hline
\endhead
1 (462) & 8 59 27.34 & -43 45 25.84 & JHK$_{\rm s}$, 1234, S, X & O9 V  &   &   & 34200$^{+1300}_{-2550}$ &  7.02$^{+ 0.14}_{- 0.14}$ & -3.32$^{+ 0.15}_{- 0.15}$ & 10.4$^{+ 0.5}_{- 0.5}$ &  8.2$^{+ 0.6}_{- 0.6}$ & 22.0$^{+ 4.0}_{- 4.0}$ \\ 
2 (408) & 8 59 28.51 & -43 46 02.94 & JHK$_{\rm s}$, 1234, S, X &  A3  IV  &  II & + &  8720$^{+ 250}_{- 260}$ &  7.40$^{+ 0.02}_{- 0.02}$ & -1.54$^{+ 0.06}_{- 0.06}$ &  8.5$^{+ 0.5}_{- 0.5}$ &  5.1$^{+ 0.5}_{- 0.5}$ &  4.9$^{+ 0.1}_{- 0.1}$ \\ 
3 (413) & 8 59 27.55 & -43 45 28.41 & JHK$_{\rm s}$, 1234, S, X & O9.7 V  &   &   & 34900$^{+1550}_{-2150}$ &  7.55$^{+ 0.01}_{- 0.01}$ & -2.71$^{+ 0.05}_{- 0.05}$ &  9.6$^{+ 0.5}_{- 0.5}$ &  6.0$^{+ 0.2}_{- 0.2}$ & 20.0$^{+ 1.0}_{- 1.0}$ \\ 
4 (292) & 8 59 21.67 & -43 45 31.05 & JHK$_{\rm s}$, 1234, S, X &  \dots &  0/I & + &  \dots &  9.32$^{+ 0.04}_{- 0.04}$ &  \dots &  8.0$^{+ 1.0}_{- 1.0}$ &  \dots &  \dots \\ 
6  & 8 59 26.24 & -43 45 27.81 & JHK$_{\rm s}$, 123, S &  G6  III  &   &   &  5450$^{+ 250}_{- 250}$ &  9.55$^{+ 0.04}_{- 0.04}$ & -1.30$^{+ 0.06}_{- 0.04}$ & 15.1$^{+ 0.1}_{- 0.4}$ &  8.7$^{+ 0.2}_{- 0.2}$ &  4.0$^{+ 0.1}_{- 0.1}$ \\ 
7  & 8 59 27.01 & -43 45 28.36 & JHK$_{\rm s}$, 12, S &  K1  IV  &   &   &  4980$^{+ 340}_{- 360}$ &  9.59$^{+ 0.01}_{- 0.01}$ & -0.61$^{+ 0.04}_{- 0.07}$ &  9.0$^{+ 0.7}_{- 0.4}$ &  6.7$^{+ 0.2}_{- 0.1}$ &  3.5$^{+ 0.1}_{- 1.1}$ \\ 
9  & 8 59 27.40 & -43 45 03.74 & JHK$_{\rm s}$, 1234, S, X &  \dots &  0/I & + &  \dots &  9.63$^{+ 0.02}_{- 0.02}$ &  \dots &  \dots &  \dots &  \dots \\ 
10 (179) & 8 59 27.74 & -43 45 38.23 & JHK$_{\rm s}$, 123, S, X &  B2.5  IV  &   &   & 19750$^{+2100}_{-1450}$ &  9.69$^{+ 0.02}_{- 0.02}$ & -0.59$^{+ 0.30}_{- 0.06}$ &  9.8$^{+ 0.5}_{- 0.5}$ &  2.9$^{+ 0.1}_{- 0.1}$ &  6.0$^{+ 1.2}_{- 0.8}$ \\ 
13  & 8 59 28.02 & -43 45 19.40 & JHK$_{\rm s}$, 1234, S &  K0  V  &  II & + &  5150$^{+ 180}_{- 190}$ & 10.08$^{+ 0.05}_{- 0.05}$ & -0.50$^{+ 0.05}_{- 0.05}$ & 12.5$^{+ 0.2}_{- 0.2}$ &  6.2$^{+ 0.2}_{- 0.2}$ &  3.3$^{+ 0.1}_{- 0.1}$ \\ 
15  & 8 59 27.86 & -43 45 59.75 & JHK$_{\rm s}$, 1234, S &  K0  V  &  II &   &  5150$^{+ 180}_{- 190}$ & 10.21$^{+ 0.05}_{- 0.05}$ & -0.30$^{+ 0.05}_{- 0.05}$ & 12.0$^{+ 0.2}_{- 0.2}$ &  5.7$^{+ 0.1}_{- 0.1}$ &  3.2$^{+ 0.0}_{- 0.1}$ \\ 
18  & 8 59 29.90 & -43 46 26.04 & JHK$_{\rm s}$, 1234, S &  G8  V  &  II & + &  5270$^{+ 270}_{- 270}$ & 10.27$^{+ 0.03}_{- 0.03}$ & -0.43$^{+ 0.03}_{- 0.03}$ & 13.7$^{+ 0.2}_{- 0.2}$ &  5.9$^{+ 0.1}_{- 0.1}$ &  3.2$^{+ 0.1}_{- 0.2}$ \\ 
26  & 8 59 24.18 & -43 45 25.91 & JHK$_{\rm s}$, 12, S, X &  K4  V  &   & + &  4490$^{+ 170}_{- 260}$ & 10.77$^{+ 0.05}_{- 0.05}$ &  0.51$^{+ 0.06}_{- 0.05}$ &  9.6$^{+ 0.3}_{- 0.4}$ &  4.3$^{+ 0.1}_{- 0.1}$ &  1.5$^{+ 0.4}_{- 0.5}$ \\ 
27  & 8 59 26.23 & -43 45 44.43 & JHK$_{\rm s}$, 1234, S &  G0  V  &  II & + &  5780$^{+ 300}_{- 300}$ & 10.79$^{+ 0.01}_{- 0.01}$ & -0.23$^{+ 0.02}_{- 0.02}$ & 16.7$^{+ 0.1}_{- 0.1}$ &  5.2$^{+ 0.0}_{- 0.0}$ &  2.7$^{+ 0.2}_{- 0.1}$ \\ 
32  & 8 59 26.74 & -43 45 30.07 & JHK$_{\rm s}$, 1234, S &  K1  IV-V  &  II & + &  4980$^{+ 340}_{- 360}$ & 11.05$^{+ 0.01}_{- 0.01}$ &  0.17$^{+ 0.04}_{- 0.07}$ & 15.3$^{+ 0.7}_{- 0.4}$ &  4.8$^{+ 0.2}_{- 0.1}$ &  2.8$^{+ 0.1}_{- 0.8}$ \\ 
34  & 8 59 28.40 & -43 44 42.84 & JHK$_{\rm s}$, 1234, S &  G3  V  &  II &   &  5580$^{+ 270}_{- 250}$ & 11.17$^{+ 0.01}_{- 0.01}$ &  0.20$^{+ 0.02}_{- 0.03}$ & 16.1$^{+ 0.3}_{- 0.2}$ &  4.3$^{+ 0.1}_{- 0.0}$ &  2.5$^{+ 0.2}_{- 0.1}$ \\ 
40  & 8 59 28.56 & -43 46 30.43 & JHK$_{\rm s}$, 123, S &  G4  IV-V  &   & + &  5550$^{+ 280}_{- 300}$ & 11.39$^{+ 0.01}_{- 0.01}$ &  0.24$^{+ 0.02}_{- 0.07}$ & 17.9$^{+ 0.7}_{- 0.1}$ &  4.3$^{+ 0.2}_{- 0.0}$ &  2.5$^{+ 0.2}_{- 0.1}$ \\ 
42  & 8 59 29.73 & -43 46 05.44 & JHK$_{\rm s}$, 123, S &  K6  V  &  II &   &  4105$^{+ 390}_{- 360}$ & 11.47$^{+ 0.01}_{- 0.01}$ &  1.13$^{+ 0.05}_{- 0.06}$ & 10.4$^{+ 0.6}_{- 0.5}$ &  3.4$^{+ 0.1}_{- 0.1}$ &  0.8$^{+ 0.6}_{- 0.3}$ \\ 
44  & 8 59 26.52 & -43 45 38.02 & JHK$_{\rm s}$, 1234, S &  K1  V  &  II & + &  4980$^{+ 190}_{- 200}$ & 11.57$^{+ 0.05}_{- 0.05}$ &  0.50$^{+ 0.06}_{- 0.06}$ & 17.1$^{+ 0.3}_{- 0.2}$ &  4.1$^{+ 0.1}_{- 0.1}$ &  2.5$^{+ 0.0}_{- 0.3}$ \\ 
47  & 8 59 27.58 & -43 45 25.00 & HK$_{\rm s}$, S &  K1  V  &   & + &  4980$^{+ 190}_{- 200}$ & 11.72$^{+ 0.03}_{- 0.03}$ &  1.29$^{+ 0.19}_{- 0.21}$ & 11.1$^{+ 1.9}_{- 1.8}$ &  2.8$^{+ 0.3}_{- 0.3}$ &  2.0$^{+ 0.1}_{- 0.1}$ \\ 
51  & 8 59 28.03 & -43 45 16.57 & JHK$_{\rm s}$, S &  K4  IV-V  &   &   &  4490$^{+ 320}_{- 390}$ & 11.79$^{+ 0.01}_{- 0.01}$ &  0.86$^{+ 0.08}_{- 0.05}$ & 15.8$^{+ 0.5}_{- 0.8}$ &  3.7$^{+ 0.1}_{- 0.1}$ &  1.4$^{+ 0.7}_{- 0.6}$ \\ 
56  & 8 59 22.63 & -43 45 15.49 & JHK$_{\rm s}$, 12, S &  K0  IV  &  II &   &  5150$^{+ 390}_{- 520}$ & 11.87$^{+ 0.01}_{- 0.01}$ &  1.67$^{+ 0.06}_{- 0.09}$ &  9.0$^{+ 0.8}_{- 0.6}$ &  2.3$^{+ 0.1}_{- 0.1}$ &  1.7$^{+ 0.1}_{- 0.2}$ \\ 
67  & 8 59 28.06 & -43 45 00.58 & JHK$_{\rm s}$, 12, S &  K3  IV  &   &   &  4630$^{+ 190}_{- 170}$ & 12.16$^{+ 0.01}_{- 0.01}$ &  1.30$^{+ 0.04}_{- 0.04}$ & 15.1$^{+ 0.4}_{- 0.4}$ &  2.9$^{+ 0.1}_{- 0.1}$ &  1.6$^{+ 0.3}_{- 0.3}$ \\ 
71  & 8 59 24.59 & -43 45 13.87 & JHK$_{\rm s}$, S &  M0  V  &   &   &  3800$^{+ 350}_{- 270}$ & 12.23$^{+ 0.01}_{- 0.01}$ &  2.15$^{+ 0.06}_{- 0.02}$ &  7.9$^{+ 0.2}_{- 0.6}$ &  2.4$^{+ 0.0}_{- 0.1}$ &  0.5$^{+ 0.3}_{- 0.2}$ \\ 
72  & 8 59 26.20 & -43 45 23.83 & JHK$_{\rm s}$, 12, S &  G5  V  &   & + &  5520$^{+ 260}_{- 350}$ & 12.25$^{+ 0.01}_{- 0.01}$ &  1.32$^{+ 0.04}_{- 0.05}$ & 15.8$^{+ 0.5}_{- 0.4}$ &  2.6$^{+ 0.1}_{- 0.1}$ &  1.8$^{+ 0.2}_{- 0.1}$ \\ 
82  & 8 59 22.61 & -43 45 49.67 & JHK$_{\rm s}$, 12, S &  K2  IV  &   &   &  4800$^{+ 360}_{- 320}$ & 12.36$^{+ 0.01}_{- 0.01}$ &  2.02$^{+ 0.05}_{- 0.08}$ & 10.4$^{+ 0.8}_{- 0.5}$ &  2.0$^{+ 0.1}_{- 0.1}$ &  1.6$^{+ 0.0}_{- 0.3}$ \\ 
83  & 8 59 25.83 & -43 45 35.83 & JHK$_{\rm s}$, 12, S &  M0  IV-V  &   &   &  3800$^{+ 350}_{- 270}$ & 12.39$^{+ 0.04}_{- 0.04}$ &  1.82$^{+ 0.07}_{- 0.05}$ & 12.5$^{+ 0.2}_{- 0.6}$ &  2.7$^{+ 0.1}_{- 0.1}$ &  0.5$^{+ 0.3}_{- 0.1}$ \\ 
85  & 8 59 29.04 & -43 45 04.50 & JHK$_{\rm s}$, 123, S &  K3  V  &  II &   &  4630$^{+ 190}_{- 170}$ & 12.45$^{+ 0.01}_{- 0.01}$ &  1.77$^{+ 0.04}_{- 0.04}$ & 13.6$^{+ 0.4}_{- 0.4}$ &  2.3$^{+ 0.0}_{- 0.0}$ &  1.5$^{+ 0.2}_{- 0.2}$ \\ 
88  & 8 59 26.87 & -43 44 47.40 & JHK$_{\rm s}$, 123, S &  K3  V  &  II & + &  4630$^{+ 190}_{- 170}$ & 12.49$^{+ 0.02}_{- 0.02}$ &  1.78$^{+ 0.04}_{- 0.04}$ & 13.8$^{+ 0.4}_{- 0.4}$ &  2.3$^{+ 0.0}_{- 0.0}$ &  1.5$^{+ 0.1}_{- 0.2}$ \\ 
93  & 8 59 29.99 & -43 45 51.88 & JHK$_{\rm s}$, 12, S &  K1  V  &  II &   &  4980$^{+ 190}_{- 200}$ & 12.55$^{+ 0.01}_{- 0.01}$ &  2.10$^{+ 0.02}_{- 0.03}$ & 11.4$^{+ 0.3}_{- 0.2}$ &  1.9$^{+ 0.0}_{- 0.0}$ &  1.5$^{+ 0.0}_{- 0.0}$ \\ 
94  & 8 59 31.21 & -43 45 52.41 & JHK$_{\rm s}$, 12, S &  K3  V  &   &   &  4630$^{+ 520}_{- 530}$ & 12.56$^{+ 0.05}_{- 0.05}$ &  1.90$^{+ 0.10}_{- 0.10}$ & 13.3$^{+ 0.8}_{- 0.8}$ &  2.2$^{+ 0.1}_{- 0.1}$ &  1.5$^{+ 0.1}_{- 0.7}$ \\ 
95  & 8 59 30.61 & -43 46 10.10 & JHK$_{\rm s}$, 12, S &  K6  V  &   &   &  4105$^{+ 170}_{- 170}$ & 12.60$^{+ 0.01}_{- 0.01}$ &  1.99$^{+ 0.02}_{- 0.03}$ & 12.9$^{+ 0.3}_{- 0.2}$ &  2.3$^{+ 0.0}_{- 0.0}$ &  0.8$^{+ 0.2}_{- 0.2}$ \\ 
97 (480) & 8 59 23.66 & -43 45 30.51 & HK$_{\rm s}$, S, X &  \dots &  0/I &   &  \dots & 12.61$^{+ 0.02}_{- 0.02}$ &  \dots &  \dots &  \dots &  \dots \\ 
100  & 8 59 24.71 & -43 45 03.32 & JHK$_{\rm s}$, 12, S &  M1  V  &   &   &  3670$^{+ 130}_{- 140}$ & 12.64$^{+ 0.01}_{- 0.01}$ &  2.40$^{+ 0.02}_{- 0.01}$ &  9.5$^{+ 0.0}_{- 0.2}$ &  2.1$^{+ 0.0}_{- 0.0}$ &  0.4$^{+ 0.1}_{- 0.1}$ \\ 
101  & 8 59 25.29 & -43 45 13.99 & JHK$_{\rm s}$, 12, S &  K1  V  &   & + &  4980$^{+ 340}_{- 360}$ & 12.68$^{+ 0.01}_{- 0.01}$ &  2.12$^{+ 0.04}_{- 0.07}$ & 12.4$^{+ 0.7}_{- 0.4}$ &  1.9$^{+ 0.1}_{- 0.0}$ &  1.5$^{+ 0.0}_{- 0.1}$ \\ 
102  & 8 59 26.46 & -43 45 53.99 & JHK$_{\rm s}$, 12, S &  M0  V  &   &   &  3800$^{+ 350}_{- 270}$ & 12.69$^{+ 0.04}_{- 0.04}$ &  1.82$^{+ 0.07}_{- 0.05}$ & 15.2$^{+ 0.2}_{- 0.6}$ &  2.9$^{+ 0.1}_{- 0.1}$ &  0.5$^{+ 0.3}_{- 0.1}$ \\ 
105  & 8 59 25.35 & -43 45 50.67 & JHK$_{\rm s}$, 123, S &  M2  V  &   & + &  3530$^{+ 270}_{- 210}$ & 12.72$^{+ 0.01}_{- 0.01}$ &  2.09$^{+ 0.02}_{- 0.07}$ & 13.1$^{+ 0.7}_{- 0.2}$ &  2.7$^{+ 0.1}_{- 0.0}$ &  0.4$^{+ 0.1}_{- 0.1}$ \\ 
108  & 8 59 27.25 & -43 46 14.62 & JHK$_{\rm s}$, 12, S &  K4  V  &   &   &  4490$^{+ 170}_{- 260}$ & 12.73$^{+ 0.02}_{- 0.02}$ &  2.75$^{+ 0.04}_{- 0.04}$ &  7.0$^{+ 0.3}_{- 0.4}$ &  1.5$^{+ 0.0}_{- 0.0}$ &  1.2$^{+ 0.0}_{- 0.2}$ \\ 
110  & 8 59 31.25 & -43 46 25.86 & JHK$_{\rm s}$, 12, S &  K3  III  &   &   &  4630$^{+ 520}_{- 530}$ & 12.77$^{+ 0.01}_{- 0.01}$ &  2.20$^{+ 0.09}_{- 0.09}$ & 12.5$^{+ 0.8}_{- 0.8}$ &  1.9$^{+ 0.1}_{- 0.1}$ &  1.4$^{+ 0.0}_{- 0.6}$ \\ 
116  & 8 59 25.08 & -43 45 45.90 & JHK$_{\rm s}$, 12, S &  M1  V  &   &   &  3670$^{+ 340}_{- 250}$ & 12.81$^{+ 0.01}_{- 0.01}$ &  1.94$^{+ 0.02}_{- 0.05}$ & 15.4$^{+ 0.5}_{- 0.1}$ &  2.7$^{+ 0.1}_{- 0.0}$ &  0.4$^{+ 0.2}_{- 0.1}$ \\ 
120  & 8 59 28.39 & -43 46 28.46 & JHK$_{\rm s}$, S &  K6  IV-V  &   &   &  4105$^{+ 390}_{- 360}$ & 12.85$^{+ 0.01}_{- 0.01}$ &  2.39$^{+ 0.05}_{- 0.06}$ & 11.4$^{+ 0.6}_{- 0.5}$ &  1.9$^{+ 0.1}_{- 0.0}$ &  0.8$^{+ 0.5}_{- 0.3}$ \\ 
124  & 8 59 31.06 & -43 45 55.54 & JHK$_{\rm s}$, 12, S &  K6  V  &  II &   &  4105$^{+ 530}_{- 490}$ & 12.91$^{+ 0.01}_{- 0.01}$ &  2.42$^{+ 0.09}_{- 0.04}$ & 11.8$^{+ 0.4}_{- 0.8}$ &  1.9$^{+ 0.0}_{- 0.1}$ &  0.8$^{+ 0.5}_{- 0.4}$ \\ 
132  & 8 59 26.99 & -43 45 52.10 & JHK$_{\rm s}$, 123, S &  K3  V  &  II &   &  4630$^{+ 360}_{- 390}$ & 12.98$^{+ 0.01}_{- 0.01}$ &  1.78$^{+ 0.07}_{- 0.07}$ & 18.4$^{+ 0.7}_{- 0.7}$ &  2.4$^{+ 0.1}_{- 0.1}$ &  1.5$^{+ 0.2}_{- 0.6}$ \\ 
134  & 8 59 28.32 & -43 45 20.85 & JHK$_{\rm s}$, S &  K3  V  &   & + &  4630$^{+ 190}_{- 170}$ & 13.00$^{+ 0.01}_{- 0.01}$ &  2.43$^{+ 0.04}_{- 0.04}$ & 12.5$^{+ 0.4}_{- 0.4}$ &  1.7$^{+ 0.0}_{- 0.0}$ &  1.3$^{+ 0.0}_{- 0.1}$ \\ 
144  & 8 59 28.12 & -43 45 55.49 & JHK$_{\rm s}$, 1, S &  K4  V  &   &   &  4490$^{+ 170}_{- 260}$ & 13.14$^{+ 0.01}_{- 0.01}$ &  2.76$^{+ 0.04}_{- 0.03}$ & 10.7$^{+ 0.3}_{- 0.4}$ &  1.5$^{+ 0.0}_{- 0.0}$ &  1.2$^{+ 0.0}_{- 0.2}$ \\ 
148  & 8 59 27.05 & -43 45 32.34 & JHK$_{\rm s}$, 123, S &  K3  IV-V  &   &   &  4630$^{+ 360}_{- 390}$ & 13.19$^{+ 0.01}_{- 0.01}$ &  2.47$^{+ 0.07}_{- 0.07}$ & 13.8$^{+ 0.7}_{- 0.7}$ &  1.7$^{+ 0.1}_{- 0.1}$ &  1.3$^{+ 0.0}_{- 0.4}$ \\ 
161  & 8 59 28.64 & -43 45 13.54 & JHK$_{\rm s}$, 12, S &  M1  IV  &  II &   &  3670$^{+ 340}_{- 250}$ & 13.26$^{+ 0.01}_{- 0.01}$ &  2.55$^{+ 0.02}_{- 0.05}$ & 13.7$^{+ 0.5}_{- 0.1}$ &  1.9$^{+ 0.0}_{- 0.0}$ &  0.4$^{+ 0.3}_{- 0.1}$ \\ 
172  & 8 59 27.91 & -43 45 04.93 & JHK$_{\rm s}$, S &  K5  IV  &   &   &  4250$^{+ 260}_{- 170}$ & 13.30$^{+ 0.01}_{- 0.01}$ &  2.36$^{+ 0.03}_{- 0.02}$ & 16.0$^{+ 0.2}_{- 0.3}$ &  1.9$^{+ 0.0}_{- 0.0}$ &  1.0$^{+ 0.3}_{- 0.2}$ \\ 
180  & 8 59 28.88 & -43 45 51.73 & JHK$_{\rm s}$, 12, S &  K6  V  &   &   &  4105$^{+ 390}_{- 360}$ & 13.39$^{+ 0.01}_{- 0.01}$ &  3.26$^{+ 0.05}_{- 0.06}$ &  8.4$^{+ 0.6}_{- 0.5}$ &  1.3$^{+ 0.0}_{- 0.0}$ &  0.9$^{+ 0.2}_{- 0.4}$ \\ 
186  & 8 59 25.30 & -43 45 31.20 & JHK$_{\rm s}$, 12, S &  K3  III  &   &   &  4630$^{+ 360}_{- 390}$ & 13.46$^{+ 0.01}_{- 0.01}$ &  2.74$^{+ 0.07}_{- 0.07}$ & 13.8$^{+ 0.7}_{- 0.7}$ &  1.5$^{+ 0.1}_{- 0.1}$ &  1.2$^{+ 0.0}_{- 0.2}$ \\ 
191  & 8 59 29.67 & -43 45 02.16 & JHK$_{\rm s}$, 12, S &  G7  V  &  II &   &  5380$^{+ 300}_{- 450}$ & 13.48$^{+ 0.05}_{- 0.05}$ &  3.15$^{+ 0.09}_{- 0.08}$ & 10.3$^{+ 0.6}_{- 0.7}$ &  1.1$^{+ 0.0}_{- 0.0}$ &  1.0$^{+ 0.1}_{- 0.0}$ \\ 

\hline
\multicolumn{13}{l}{$^a$: Id from \citet{Bik2005, Bik2006, Ellerbroek2013}. $^b$: Available data: SOFI (JHK$_{\rm s}$), IRAC (bands 1234), SINFONI (S), X-shooter (X).}\\
\multicolumn{13}{l}{$^c$: Based on comparison with reference spectra from \cite{Rayner2009}. For objects 1, 3 and 10, see Sec.~\ref{sec:spectroscopy:et}.}\\
\multicolumn{13}{l}{$^d$: Evolutionary class \citep{Lada1987} derived from IRAC colors as defined by \citet{Megeath2004, Gutermuth2009}.}\\
\multicolumn{13}{l}{$^e$: From \citet{Kenyon1995}, averaged between the spectral type range. For objects 1, 3 and 10, see Sec.~\ref{sec:spectroscopy:et}.}\\
\multicolumn{13}{l}{$^f$: Assuming a distance of 0.7~kpc. $^g$: Derived using intrinsic colors of \citet{Kenyon1995} and \citet{Martins2006}. For objects 1, 2, 3 and 10, SED-fitting was used (Sect.~\ref{sec:spectroscopy:et}).} \\
\multicolumn{13}{l}{$^h$: Result of fitting with Kurucz model \citep{Kurucz1993} $^i$: Result of interpolating evolutionary tracks and isochrones \citep{Siess2000, DaRio2009, Ekstrom2012}.}\\
\label{tab:dataoverview}
\end{longtable} 
}

\twocolumn

\clearpage

\begin{figure*}[!h]
\centering
\includegraphics[width=0.95\textwidth, page=1]{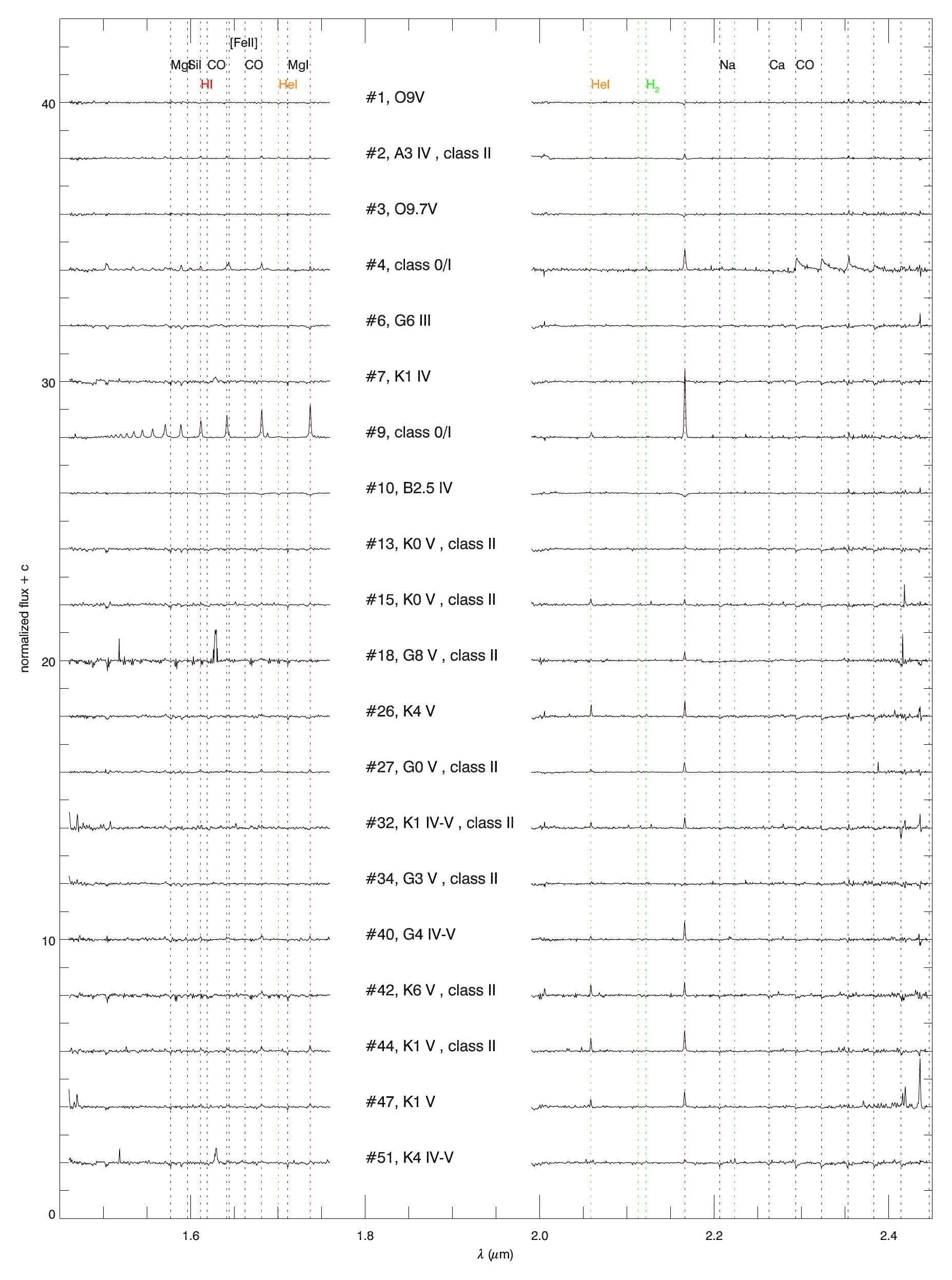}
\caption{\label{fig:sinfonispectra} SINFONI spectra; spectral lines used for classification are indicated.}
\end{figure*}
\addtocounter{figure}{-1}

\begin{figure*}[!h]
\centering
\includegraphics[width=0.95\textwidth, page=2]{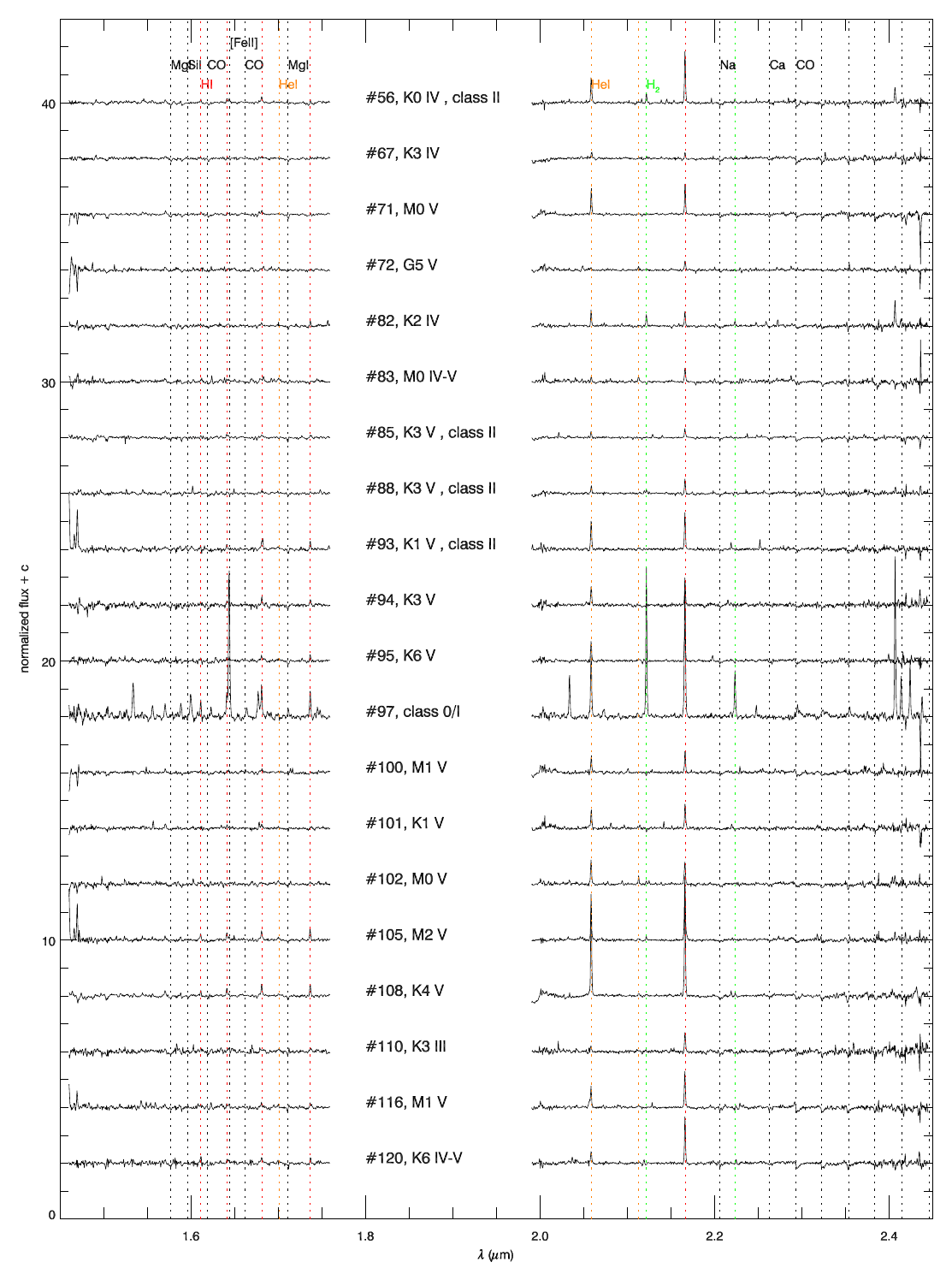}
\caption{Continued.}
\end{figure*}
\addtocounter{figure}{-1}

\begin{figure*}[!h]
\centering
\includegraphics[width=0.95\textwidth, page=3]{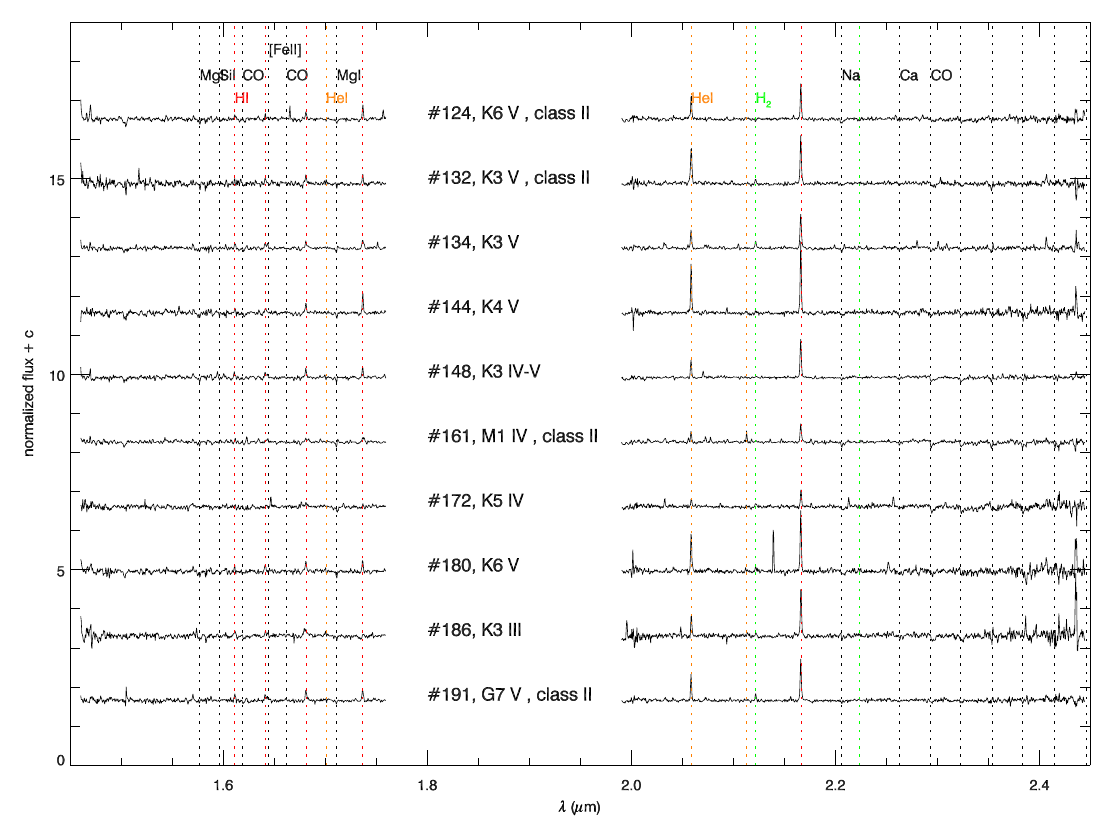}
\caption{Continued.}
\end{figure*}

\end{document}